\documentclass[prd,byrevtex,showpacs,showkeys%
,nofootinbib
,preprint
]{revtex4}
\usepackage{amssymb,amsmath,amsxtra}
\usepackage[dvipdfm]{graphicx}
\usepackage{float}
\newcommand{\pa}{\partial}
\newcommand{\del}{\delta}
\newcommand{\tphi}{\tilde{\phi}}
\newcommand{\om}{\omega}

\newcommand{\ta}{\tilde{a}}
\begin{document}
\title{Simulation of Acoustic Black Hole in a Laval Nozzle}
\author{Hironobu Furuhashi}
\email{hironobu@gravity.phys.nagoya-u.ac.jp} \author{Yasusada Nambu}
\email{nambu@gravity.phys.nagoya-u.ac.jp} \affiliation{Department of
  Physics, Graduate School of Science, Nagoya University, Nagoya
  464-8602, Japan}
\author{Hiromi Saida}
\email{saida@daido-it.ac.jp}
\affiliation{Department of Physics, Daido Institute of Technology,
  Nagoya 457-8530, Japan}
\date{March 24, 2006}
\begin{abstract}
  A numerical simulation of fluid flows in a Laval nozzle is performed
  to observe the formation of an acoustic black hole and the
  classical counterpart to Hawking radiation under a realistic
  setting of the laboratory experiment. We aim to
  construct a practical procedure of the data analysis to extract
  the classical counterpart to Hawking radiation from experimental data.
  Following our procedure, we determine the surface gravity of the
  acoustic black hole from obtained numerical data. Some noteworthy
  points in analyzing the experimental data are clarified through our
  numerical simulation.
\end{abstract}

\maketitle
\section{\label{sec:intro}Introduction}

The existence of the black hole horizon is essential to the Hawking
radiation
\cite{HawkingSW:CMP43:1975,BirrellND:CUP:1982,VisserM:IJMPD12:2003,UnruhW:PRD71:2005}.
The horizon causes an extremely large redshift on an outgoing wave of a
matter field during propagating from a vicinity of the horizon to
the asymptotically flat region. This redshift results in the Planckian
distribution of quantum mechanically created particles at the
future null infinity. Although the Hawking radiation has not yet been
confirmed observationally, it has been pointed out that the similar
phenomenon appears on phonons in a transonic flow of a fluid. This
theoretical phenomenon in the fluid is called the sonic analogue of
Hawking radiation, and the transonic fluid flow on which the sonic
analogue of Hawking radiation occurs is called the acoustic black hole
\cite{UnruhWG:PRL46,UnruhWG:PRD51:1995,VisserM:CQG15:1998,VolovikGE:JETPL69:1999,%
NovelloM:WS:2002,BarceloC:INJMPA18:2003,BarceloC:NJP6:2004,BarceloC:0505065:2005}.
In the transonic flow, the boundary between a subsonic and a supersonic
region is the sonic point. The sonic point corresponds to the black
hole horizon and called the sonic horizon. The region apart form
the sonic point in the subsonic region corresponds to the asymptotically
flat region in a black hole spacetime. The sound wave which propagates
from a vicinity of the sonic point to the subsonic region receives
the extremely large redshift and the sonic analogue of Hawking radiation
appears on phonons. The Hawking temperature $T_H$ of the acoustic black
hole is given by the gradient of the fluid velocity at the sonic point
$x=x_s$ \cite{UnruhWG:PRL46}:
\begin{equation}
   T_H = \frac{\hbar\, c_s}{2 \pi k_B} 
\left. \frac{d}{dx}\left(\frac{v}{c_s}\right)\right|_{x = x_s} \, ,
\label{eq:H-temp}
\end{equation}
where $c_s$ and $v$ are the sound velocity and the fluid velocity,
respectively. For an ordinary system with typical size $\sim 1$ m and
$c_s \sim 340$ m/s, we obtain $T_H \sim 4 \times 10^{-10}$ K. Thus, in
a practical experiment in a laboratory, it is very difficult to detect
the sonic analogue of Hawking radiation with such an extremely low
temperature.

While it is very hard to detect the sonic analogue of Hawking
radiation on phonons, which is a quantum origin, a classical sound
wave with a sufficiently large amplitude can be available for detecting
the effect of the acoustic black hole on a propagation of sound waves
in practical experiments. Indeed, the thermal nature of Hawking
radiation is not a quantum origin; it comes from a large redshift on
the wave propagating from a vicinity of the horizon to the asymptotically
flat region, and has nothing to do with the quantum effect (see Appendix).
Therefore, if we can construct a classical quantity from sound waves
which shows the thermal distribution, then such a quantity can be
interpreted as a classical counterpart to Hawking radiation. A
candidate for the classical counterpart has already been introduced
in the references \cite{NouriZonozM:9812088:1998,SakagamiM:PTP107:2002}
by using a Fourier component of an outgoing wave from a vicinity
of the horizon.

By the way, since a realistic fluid is composed of molecules, the effect
of the molecule size discrepancy of the fluid becomes one of the important
issues of the acoustic black hole. This effect appears in the modified
dispersion relation at high energies of phonons. However, in the expansion
by the molecule size discrepancy, the zeroth order form of the quantum
expectation value of the number operator $\left< 0 \right| N \left| 0 \right>$
of phonons results in the thermal spectrum at least not so high energy
region. That is, the effect of the molecule size discrepancy is not dominant
in the sonic analogue of Hawking radiation. Therefore, before proceeding to
the issue of the molecule size discrepancy, it is necessary to detect the
ordinary Hawking radiation (thermal spectrum) by a real experiment in
laboratory, and the molecule size discrepancy is not in the scope of this
paper. Hence we concentrate on the construction of the practical procedure
to extract the evidence of the Hawking radiation (thermal spectrum) from
the experimental data. To do so, we make use of the classical counterpart
to Hawking radiation. 

We should emphasize here that the Hawking temperature (\ref{eq:H-temp})
includes the Planck constant $\hbar$ and we can not consider the
temperature $T_H$ in discussing the ``classical'' counterpart to Hawking
radiation. However we can consider the surface gravity $\kappa$ of the
black hole horizon which is a purely classical quantity and relates to
the Hawking temperature as $k_B \, T_H = \hbar \, \kappa/2 \pi$. Hence,
for the acoustic black hole, we consider the surface gravity $\kappa_H$
of the sonic horizon instead of the Hawking temperature,
\begin{equation}
 \kappa_H
  = c_s \left.
    \frac{d}{dx}\left(\frac{v}{c_s}\right)\right|_{x = x_s} \, .
\label{eq:surf.grav}
\end{equation}

In this paper, we aim to construct a practical procedure of the data
analysis to obtain the classical counterpart to Hawking radiation in
a realistic setting of the acoustic black hole. To do so, we perform
a numerical simulation of a fluid flow in a practical experimental
setting, and demonstrate that our procedure works well to detect the
classical counterpart to Hawking radiation.

We organize the paper as follows. In the section \ref{sec:counterpart},
we review the acoustic black hole in a Laval nozzle and define the
classical counterpart to Hawking radiation. The section
\ref{sec:numerical} is devoted to the introduction of our numerical
method to simulate a transonic flow, and to the construction of the
practical procedure of the data analysis. In the section \ref{sec:result},
our numerical results are presented and we determine the surface gravity
of the sonic horizon. Finally the summary and conclusion are in
the section \ref{sec:summary}.

\section{Acoustic black hole with a Laval nozzle\label{sec:counterpart}}

\begin{figure}[H]
\centering
\includegraphics[width=0.3\linewidth,clip]{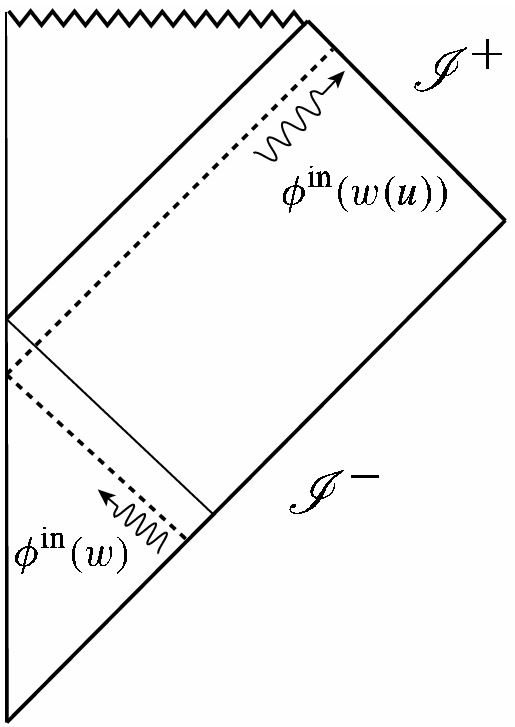} \hspace{10mm}
\includegraphics[width=0.4\linewidth,clip]{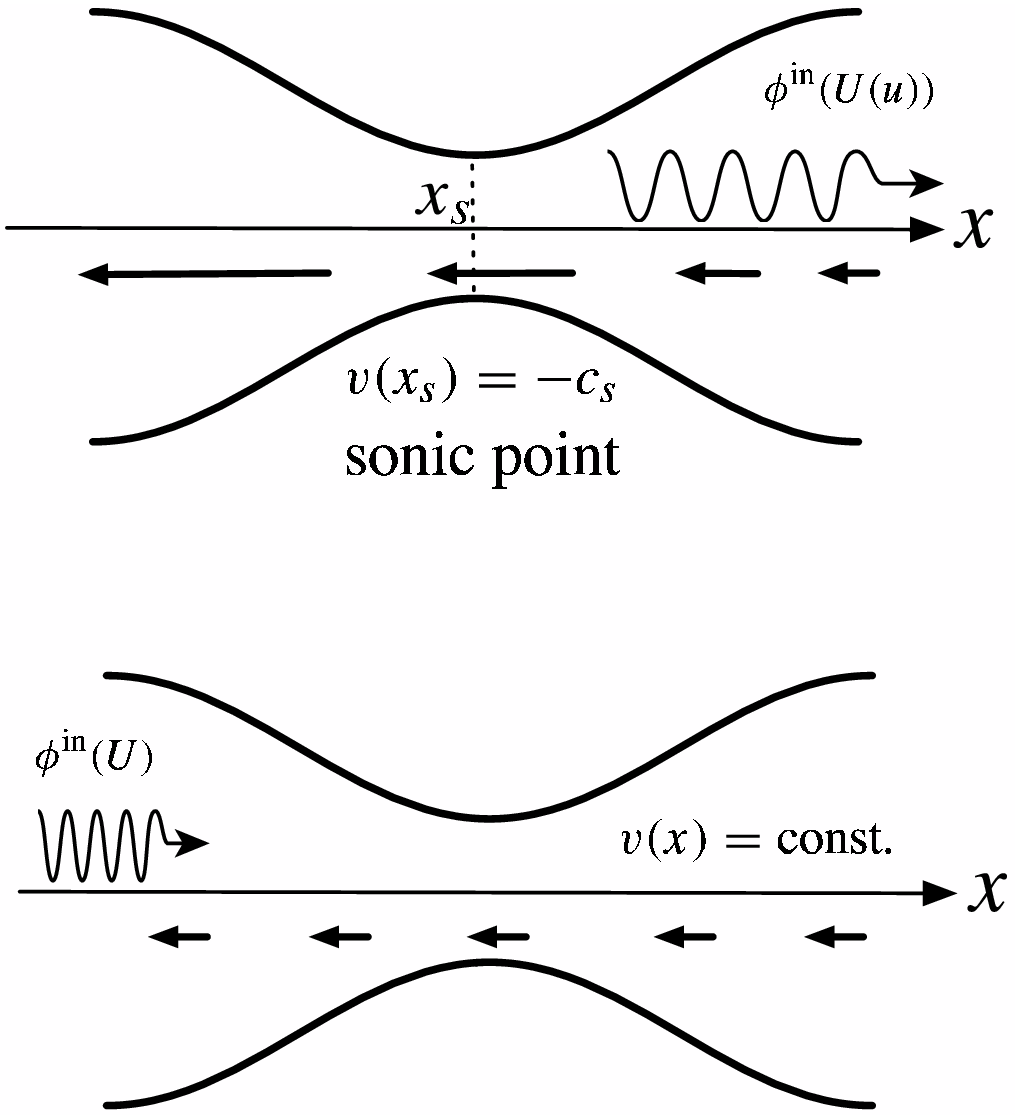}
\caption{Propagation of waves in a spacetime of gravitational collapse and
  in a Laval nozzle.}
\label{fig:nozzle}
\end{figure}

We consider an acoustic black hole with a fluid in a Laval
nozzle as proposed in \cite{SakagamiM:PTP107:2002}. The Laval nozzle,
as shown in the right panel in FIG.~\ref{fig:nozzle}, has a
throat where the cross section of the nozzle becomes minimum. The
fluid is accelerated from the up stream to the down stream and the flow
has the sonic point at the throat when a sufficiently large velocity is
given at the inlet of the nozzle. Even if we prepare an initial flow
which has no sonic point (e.g. the lower part of the right panel in
FIG.~\ref{fig:nozzle}), the flow can settle down to a stationary
transonic flow (e.g. the upper part of the right panel in FIG.~
\ref{fig:nozzle}) with appropriate boundary conditions at the inlet
and the outlet of the nozzle. This is the sonic analogue model which
corresponds to the gravitational collapse.

Here we note that, if a stationary transonic flow is prepared from the
beginning (which corresponds to the eternal black hole), we can not
expect to obtain the classical counterpart to the Hawking radiation;
for the eternal case, the tunneling of phonons across the
sonic point can result in the Hawking radiation and this is the purely
quantum effect.  Hence, when we are interested in the ``classical''
counterpart to Hawking radiation, it is necessary to consider the
situation of the sonic point formation in course of the dynamical
 evolution of the fluid flow . As seen below, the sound wave
which is prepared on the fluid flow before the formation of the sonic
point becomes the classical counterpart to the quantum fluctuation
which causes the classical counterpart to Hawking radiation.

\subsection{Basic equations\label{sec:counterpart:basic-eq}}
\begin{figure}[H]
  \centering
  \includegraphics[width=0.4\linewidth,clip]{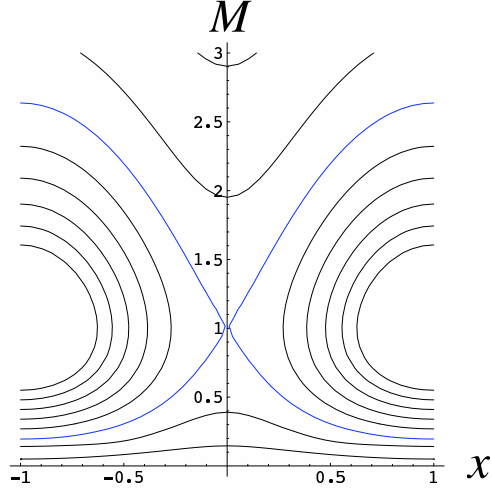}
  \caption{Absolute value of the Mach number as a function of a spatial
    coordinate $x$. Each line represents the stationary solution of the
    flow with different boundary value $M_{in}$. The blue lines represent
    transonic flows.}
  \label{fig:mach}
\end{figure}

We consider a perfect fluid and treat the flow in a Laval nozzle as
one dimensional for simplicity. The  basic equations are the mass
conservation equation and the Euler equation:
\begin{subequations}
\label{eq:fluid}
\begin{align}
&\pa_t \rho + \frac{1}{S} \partial_x\left(\rho v S \right) = 0, \label{eq:fluidA}\\
&\rho\left(\pa_t v + v\pa_x v \right) = - \pa_x p, \label{eq:fluidB}
\end{align}
\end{subequations}
where $\rho(t,x)$ is the mass density, $v(t,x)$ is the fluid velocity,
$p(t,x)$ is the pressure of the fluid and $S(x)$ is the cross section
of the Laval nozzle. We assume the adiabatic ideal gas type equation
of state $p \propto \rho^{\gamma}$ where $\gamma$ is the adiabatic
index. The sound velocity $c_s$ is given by
\begin{equation}
 c_s^2 \equiv
\dfrac{dp}{d\rho} \propto \rho^{\gamma - 1}.
\end{equation}
For a stationary background flow, we can obtain the sound velocity
$c_s$ and the cross section $S$ of the nozzle as a function of the Mach
number $M=v/c_s$ of the flow:
\begin{equation}
 c_s=c_{{\text{in}}}\left(\frac{M^2+\frac{2}{\gamma-1}}{M_{\text{in}}^2+\frac{2}{\gamma-1}}
\right)^{-1/2},\quad
  S=S_{\text{in}}\frac{M_{\text{in}}}{M}\left(\frac{M^2+\frac{2}{\gamma-1}}{M_{\text{in}}^2+
    \frac{2}{\gamma-1}}\right)^{\frac{\gamma+1}{2(\gamma-1)}},
\label{eq:stationary}
\end{equation}
where quantities with subscript ``in'' represent the values at the inlet
of the nozzle. The spatial derivative of the Mach number at the sonic
point $x = x_s$ is
\begin{equation}
  \left.\frac{dM}{dx}\right|_{x_s}
=\pm\left.\sqrt{\frac{(\gamma+1)}{4}\frac{1}{S}\left(\frac{d^2S}{dx^2}\right)}
 \right|_{x_s},
\end{equation}
and this value relates to the surface gravity of the sonic horizon by
Eq.~\eqref{eq:surf.grav}. Figure~\ref{fig:mach} shows the structure of
the stationary flow for $\gamma=7/5$ and $S(x)=2-\cos(\pi x)$. Here we set
that the inlet is at $x=1$, the throat at $x=0$ and the outlet at $x=-1$.
That is, the fluid flows downward from $x=1$ to $x=-1$, then $M<0$ and 
FIG.~\ref{fig:mach} shows the absolute value of $M$. The boundary
condition is given by the Mach number $M_{\text{in}}$ at the inlet of
the nozzle, and each line in FIG.~\ref{fig:mach} corresponds to
a different value of $M_{\text{in}}$. The blue line which passes
the sonic point $(x,M)=(0,1)$ is given by
$M_{\text{in}}= M_*\approx - 0.197$ and represents the stationary
transonic flow. For the case $c_{\text{in}}=1$, we find
\begin{equation}
  \left.\frac{dM}{dx}\right|_{x=0}=\pm \sqrt{\frac{3}{5}}\,\pi,\quad
  c_s|_{x=0}=(3|M_*|)^{1/6}, 
\end{equation}
and the surface gravity given by Eq.~\eqref{eq:surf.grav} is
\begin{equation}
  \frac{\kappa_H}{2\pi}
  =\frac{(3|M_*|)^{1/6}}{2\pi}\sqrt{\frac{3}{5}}\,\pi \approx 0.355.
  \label{eq:temp-th}
\end{equation}
Here it should be emphasized that, because the cross section of the nozzle is
set $S(x)=2-\cos(\pi x)$ with $-1 < x < 1$, the spatial scale is normalized
by $L/2$, where $L$ is the length of the nozzle, and that the temporal
scale is normalized by $L/(2c_{\text{in}})$, where we set $c_{\text{in}} = 1$. 

We have not considered the perturbation of the fluid flow so far. In the
following sections, the perturbation (sound wave) is introduced on the
fluid flow, and the classical counterpart to Hawking radiation is defined.
Then in the section~\ref{sec:result}, we compare this surface gravity
(\ref{eq:temp-th}) with the surface gravity of the sonic horizon which is
obtained through the observation of the classical counterpart to Hawking
radiation in our numerical simulation.

\subsection{Classical counterpart to  Hawking radiation}

As originally shown in the reference \cite{UnruhWG:PRL46}, the
perturbation of the velocity potential for the transonic fluid flow
obeys the same equation as a massless free scalar field in a black
hole spacetime. Therefore the classical counterpart to Hawking
radiation of the acoustic black hole in the Laval nozzle should be
observed by the sound wave on the transonic fluid flow. The most
important sound wave is the one which starts to propagate against
the stream from the outlet of the nozzle before the formation of
the sonic point (the lower part of the right panel in FIG.
~\ref{fig:nozzle}) and passes through the throat just before
the moment of the sonic point formation to reach the inlet of
the nozzle (the upper part of the right panel in FIG.~\ref{fig:nozzle}).
This sound wave receives the extremely large redshift which will cause
the classical counterpart to Hawking radiation.

When the fluid flow is irrotational and we introduce the velocity
potential $\Phi$ which relates with the fluid velocity as
$v = \partial_x \Phi$, the evolution equation of the perturbation
$\phi$ of $\Phi$ is obtained from Eqs.~\eqref{eq:fluid},
\begin{equation}
  \label{eq:wave-eq}
  \frac{1}{S\,\rho}\left[\pa_t+\pa_x
    v\right]\left(\frac{S\,\rho}{c_s^2}\right)\left[\pa_t+v\pa_x\right]\phi
    -\frac{1}{S\,\rho}\pa_x\!\left(S\,\rho\pa_x\phi\right)=0.
\end{equation}
This corresponds to the Klein-Gordon equation $\square\phi=0$ with
the acoustic metric
\begin{equation}
  \label{eq:metric1}
  ds^2=\left(\frac{S\,\rho}{c_s}\right)
       \left[-(c_s^2-v^2)dt^2-2v\,dt\,dx+dx^2+dy^2+dz^2\right],
\end{equation}
where the location of the sonic horizon (sonic point) $x=x_s$ is given by
$c_s(x_s)^2=v(x_s)^2$. Here we should note that, because we consider the
one dimensional flow along $x$-axis as mentioned at the beginning of the
previous section \ref{sec:counterpart:basic-eq}, the two dimensional
section of $(y,z)$ coordinate is omitted in obtaining
Eq.~\eqref{eq:wave-eq} from $\square\phi=0$. In the following analysis,
we omit the two dimensional section of
$(y,z)$ coordinate.

In order to describe the extremely large redshift which causes
the classical counterpart to Hawking radiation, we introduce
the following null coordinates for the first,
\begin{equation}
  u=t-\int\frac{dx}{c_s+v},\quad w=t+\int\frac{dx}{c_s-v},
  \label{eq:null1}
\end{equation}
and the acoustic metric becomes
$ds^2 = \left(S\,\rho/c_s\right)\left(c_s^2-v^2\right)du\,dw$, which
has a coordinate singularity at $x=x_s$. In order to eliminate
the coordinate singularity, we introduce the new null coordinates
\begin{equation}
  U=-\frac{1}{c_{s0}'+v_0'}\exp\left[-(c_{s0}'+v_0')u\right],\quad 
  W=\frac{1}{c_{s0}'+v_0'}\exp\left[(c_{s0}'+v_0')w\right],
\end{equation}
where $'=\pa_x$ and quantities with subscript $0$ denotes values at
the sonic point $x=x_s$. Then the acoustic metric and the evolution
equation \eqref{eq:wave-eq} of the perturbation $\phi$ become
\begin{align}
  ds^2&=-\left(\frac{S\,\rho}{c_s}\right)\,(c_s^2-v^2)\,
        \exp\left[-2(c_{s0}'+v_0')\int\frac{c_s\,dx}{c_s^2-v^2}\right]\,dU\,dW
       \equiv -F\,dU\,dW, \label{eq:metric2} \\
  \square\phi&=\frac{1}{F}\pa_{UW}\,\phi=0.
\end{align}
The form of this metric near the sonic horizon is
$ds^2\approx-2S_0\rho_0(c_{s0}'+v_0')\,dU\,dW$, where 
$c_s^2-v^2\approx 2c_{s0}(c_{s0}'+v_0')x$ is used. This denotes
explicitly that the coordinate $(U,W)$ does not have coordinate
singularities and corresponds to the Kruskal-Szekeres coordinate of the
Schwarzschild spacetime. Here it should be recalled that, in the spacetime
of the gravitational collapse (the left panel in FIG.~\ref{fig:nozzle}),
the Kruskal-Szekeres coordinate is appropriate to describe the rest
observer at the spacetime region before the formation of the black hole
\cite{BirrellND:CUP:1982}. Hence the outgoing normal mode of the sound
wave which corresponds to the zero point fluctuation of the quantized 
matter field before the formation of the black hole (i.e. the mode
function on the flat spacetime) is given by
\begin{equation}
  \phi^{\text{in}}(U)=A_{\phi}\,e^{-i\omega_0U},
\label{eq:input}
\end{equation}
where $\om_0$ is the initial frequency and $A_{\phi}$ is the amplitude
of this mode. In the original coordinate $(t,x)$, this mode behaves as
\begin{equation}
  \phi^{\text{in}}(U(u))=\phi^{\text{in}}(t,x)
  =A_{\phi}\exp\left[\frac{i\omega_0}{c_{s0}'+v_0'}
\exp\left(-(c_{s0}'+v_0')\left(t-\int^x\frac{dx}{c_s+v}\right)\right)\right].
  \end{equation}
This yields the temporal wave form at the observation point
$x=x_{\text{obs}}$ in the asymptotic region:
\begin{equation}
  \phi^{\text{out}}(t)\equiv \phi^{\text{in}}(t,x_{\text{obs}})
 =A_{\phi}\exp\left[\frac{i\omega_0}{c_{s0}'+v_0'}
 \exp\left(-(c_{s0}'+v_0')\left(t-\int^{x_{\text{obs}}}\frac{dx}{c_s+v}\right)\right)\right].
\label{eq:output}
\end{equation}
Fourier component of this mode is defined by
\begin{equation}
  \phi_{\omega}^{\text{out}}=\int_{-\infty}^{\infty}dt\,\phi^{\text{out}}(t)e^{i\omega t},
\label{eq:fourier}
\end{equation}
and the power spectrum is
\begin{equation}
  \label{eq:planck}
  P(\omega)=\left|\phi_{\omega}^{\text{out}}\right|^2=
  \left(\frac{A_{\phi}^2}{\om_H\,\omega}\right)\frac{e^{\omega/\om_H}}{e^{\omega/\om_H}-1}
  \qquad(-\infty<\omega <+\infty),
\end{equation}
where the quantity $\om_H$ is defined by
\begin{equation}
  \om_H
 \equiv\frac{\kappa_H}{2\pi}
 =\frac{c_{s0}}{2\pi}\frac{d}{dx}\!\left(\frac{v}{c_s}\right)_0.
\end{equation}
Here it should be noted that, because the existence of the sonic point
is assumed in the above discussion, the form of $P(\omega)$ of
Eq.~\eqref{eq:planck} must be obtained after the formation of the sonic
horizon in the fluid flow.

One may think it is strange that the initial frequency $\om_0$
does not appear in the power spectrum \eqref{eq:planck}. It is the fact
that the initial frequency $\om_0$ does appear in the power spectrum in
the context of the quantum field theory, because the amplitude $A_{\phi}$
is determined by the normalization condition for the mode functions to
be $A_{\phi} \propto 1/\sqrt{\om_0}$. However, in the calculation for
the classical counterpart to Hawking radiation, the initial frequency
$\om_0$ appears only in the phase of the Fourier component
$\phi_{\om}^{\text{out}}$ and the power $P(\om)$ does not explicitly
depend on $\om_0$.

The power spectrum \eqref{eq:planck} has the Planckian distribution for
$\om<0$ and this is the classical counterpart to the quantum Hawking
radiation \cite{NouriZonozM:9812088:1998,SakagamiM:PTP107:2002}. For the
quantum Hawking radiation, the magnitude of the power is determined by
the zero point oscillation of the quantized field. However, for the
classical counterpart to Hawking radiation, the magnitude of the power
depends on the amplitude $A_\phi$ of the input signal and it is expected
that we can detect the thermal distribution of the power spectrum for the
sound wave of a sufficiently large amplitude in practical experiments.

As long as concerning one dimensional fluid flow, the amplitude of the
sound wave does not decrease during propagating from the outlet to
the inlet of the nozzle. The one dimensional transonic fluid flow
corresponds to the two dimensional black hole spacetime in which no
curvature scattering occurs. The theoretically expected power
\eqref{eq:planck} is equivalent to the formula of the Hawking radiation
for black holes derived ignoring  the curvature scattering.

Our purpose is to observe this thermal spectrum via the numerical
simulation of a transonic flow in the Laval nozzle, and to construct the
practical procedure of the data analysis. Because the observable sound
wave is a real valued one, we prepare the following real valued input mode
at the outlet of the nozzle before the formation of the sonic horizon:
\begin{equation}
   \phi^{\text{in}(\theta)}(t)
 = A_{\phi}\cos\left(\om_0\,t+\theta\right)
 = \frac{1}{2}
   \left[e^{-i\theta}\phi^{\text{in}}(U(t,x_{\text{e}}))
       + e^{i\theta}\phi^{\text{in}*}(U(t,x_{\text{e}}))\right],
\label{eq:real-input}
\end{equation}
where $\theta$ is the initial phase of the wave, $\phi^{\text{in}}(U)$ is given by
Eq.~\eqref{eq:input}, and $x=x_{\text{e}}$ is the location of the outlet
of the nozzle. Here we should note that, although the outlet of the
nozzle is the ``exit'' of the background fluid flow, the sound wave can
propagate against the background flow from the outlet to the inlet of
the nozzle. Then, since the amplitude of the sound wave on one dimensional
fluid flow does not decrease, the input mode of Eq.\eqref{eq:real-input}
causes the following sound wave in the fluid flow,
\begin{equation}
 \phi^{(\theta)}(t,x) =
 \frac{1}{2}
 \left[e^{-i\theta}\phi^{\text{in}}(U(t,x))
     + e^{i\theta}\phi^{\text{in}*}(U(t,x))\right].
\end{equation}
The effect of the redshift due to the formation of the acoustic black
hole gradually appears on the observed sound wave. The temporal
evolution of the observed sound wave is affected by the formation of
the acoustic black hole. That is, the information of the classical
counterpart to Hawking radiation is encoded in the observed sound wave.
Therefore, using the output form \eqref{eq:output}, the resulting output
signal after the formation of the sonic horizon is given by
\begin{equation}
  \phi^{\text{out}(\theta)}(t)=
  \frac{1}{2}\left[e^{-i\theta}\phi^{\text{out}}(t)+
    e^{i\theta}\phi^{\text{out}*}(t)\right],
\end{equation}
and referring Eq.~\eqref{eq:fourier}, the Fourier component of this output
signal is given by
\begin{equation}
  \phi^{\text{out}(\theta)}_{\omega}=\frac{1}{2}\left(e^{-i\theta}\phi^{\text{out}}_{\omega}
+e^{i\theta}\phi_{\omega}^{\text{out}*}\right).
\label{eq:real-output}
\end{equation}
Hence, because the initial phase $\theta$ appears in
$\phi^{\text{out}(\theta)}_{\omega}$, we can not obtain the pure
Fourier component $\phi_{\omega}^{\text{out}}$ by observing once the
real valued output signal $\phi^{\text{out}(\theta)}(t)$. In order to
retrieve the pure Fourier component $\phi^{\text{out}}_{\omega}$ from
the mixed Fourier component $\phi^{\text{out}(\theta)}_{\omega}$, we
have to superpose two output modes $\phi^{\text{out}(\theta)}$ and
$\phi^{\text{out}(\theta+\pi/2)}$, and obtain
\begin{equation}
  \phi_{\omega}^{\text{out}}=e^{i\theta}\left(\phi_{\omega}^{\text{out}(\theta)}
    +i\phi_{\omega}^{\text{out}(\theta+\pi/2)}\right).
  \label{eq:superpose}
\end{equation}
It is obvious that the power spectrum of this Fourier component gives
the same form as Eq.~\eqref{eq:planck}. Therefore, in order to obtain
the Planckian distribution in the power spectrum $P(\om)$ of the observed
sound wave from real valued input signals, we have to combine at least
two output modes of which the input phases differ by $\pi/2$.

\section{\label{sec:numerical}%
Numerical simulation of transonic flows in a Laval nozzle}
Our numerical simulation is designed to generate a one dimensional
transonic fluid flow from the initial configuration with no sonic point.
We prepare the input signal at the outlet of the nozzle and observe
the sound wave at the inlet of the nozzle. At the same time,
the practical procedure of the data analysis is constructed.

\subsection{Basic equations for numerical simulations}
For the numerical calculation, we rewrite the fluid Eqs.~\eqref{eq:fluid}
 using $c_s$ and $v$, 
\begin{subequations}
\begin{align}
& \frac{2}{\gamma-1}\pa_t c_s+\frac{2}{\gamma-1}v\pa_x c_s+c_s\pa_x v +
\left(\frac{\pa_x S}{S}\right)c_s v =0, \\
&\pa_t v+v\pa_x v+\frac{2}{\gamma-1}c_s\pa_x c_s = 0.
\end{align}
\end{subequations}
This set of equations is not suitable for a numerical simulation of
wave propagation in a transonic flow. We transform these
equations to the advection form. We introduce the Riemann invariants
$J_{\pm}(t,x)$ as follows 
\begin{equation}
 J_{\pm} \equiv v \pm
 \frac{2c_s}{\gamma -1}.
 \label{eq:riemann}
\end{equation}
Then the basic equations become
\begin{subequations}
  \label{eq:num}
\begin{align}
 &\pa_t J_+ +V_{+}\pa_x J_{+} =
 -\frac{\gamma-1}{8}\left(\frac{\pa_xS}{S}\right)\left(J_{+}^2-J_{-}^2\right), 
  \label{eq:num1}\\
 &\pa_t J_{-}-V_{-}\pa_x J_{-} =+\frac{\gamma-1}{8}\left(\frac{\pa_x
     S}{S}\right)\left(J_+^2-J_-^2\right), \label{eq:num2}
\end{align}
\end{subequations}
 where
\begin{equation}
 V_{+} = c_s + v = \frac{\gamma + 1}{4}J_+-\frac{\gamma-3}{4}J_-,
 \qquad 
 V_{-} = c_s - v =\frac{\gamma-3}{4}J_+-\frac{\gamma + 1}{4}J_-.
\end{equation}
The left hand side of Eqs.~\eqref{eq:num} are of the advection
form. That is, if $\partial_x S = 0$ and $\partial_tV_{\pm}=0$, $J_+$
propagates upward along $t-\int dx/V_{+}=$ const. and $J_{-}$
propagates downward along $t+\int dx/V_{-}=$ const. This form of the
equations is suitable to treat the propagation of waves numerically.

In the sonic analogue of Hawking radiation, the perturbation of the
velocity potential (sound wave) corresponds to the scalar field in
a black hole spacetime as shown by Eq.~\eqref{eq:wave-eq}. Therefore,
in a practical experiment of an acoustic black hole, we should calculate
the velocity potential by integrating the observed velocity of the fluid:
\begin{equation}
 \Phi(t,x_{\text{obs}})
\equiv \int_{x_0}^{x_{\text{obs}}} \, dx\,v(t,x) \, , 
\end{equation}
where $x_{\text{obs}}$ is the observation point of sound waves and
$x_0$ defines the origin of the velocity potential. However, since our
experimental setting is designed to observe the sound wave at a fixed
spatial point $x = x_{\text{obs}}$, the observational data is a
temporal sequence of $v(t,x_{\text{obs}})$ and we must devise the
other method to obtain the velocity potential at the observation
point. In the upstream subsonic region of the transonic flow where
the effects of the sonic horizon is negligible and the flow is
stationary, the sound wave propagates along the ``null'' direction as 
indicated by Eq.~\eqref{eq:null1}
\begin{equation}
t-\frac{x}{\left.(c_s+v)\right|_{x_{\text{obs}}}}=\text{const.}
\end{equation}
and we can obtain the value of the velocity potential at $x_{\text{obs}}$
by integrating the velocity with respect to time
\begin{equation}
 \Phi(t,x_{\text{obs}})=\left.(c_s+v)\right|_{x_{\text{obs}}}
 \int_{t_0}^t dt \,v(t,x_{\text{obs}}),
 \label{eq:potential}
\end{equation}
where $t_0$ defines the origin of the potential. We use this formula to
evaluate the velocity potential at the observation point.

\subsection{Experimental setting and numerical method}

We use the Laval nozzle with the cross section of the following form,
\begin{equation}
 S(x) = 2 - \cos\left(\pi x \right)  \qquad (-1 \leq x \leq 1) .  
\label{eq:cross-section}
\end{equation}
We make the fluid flow against $x$-axis and assume $v < 0$. The inlet of
the nozzle is at $x = 1$, the throat at $x = 0$ and the outlet at
$x = -1$. The input wave is prepared at the outlet of the nozzle and
the wave propagates against the flow from $x = -1$ to $x = 1$. After the
flow settles down to a stationary transonic flow, since
$\left.\partial_x S \right|_{x=1} = 0$ at the inlet, the set of
Eqs.~\eqref{eq:num} gives $\left.\pa_x v \right|_{x=1} = 0$ at the inlet.
This means that the inlet of the nozzle corresponds to the asymptotically
flat region in the gravitational collapse spacetime. Hence we make
the observation of the sound wave at the inlet and set
$x_{\text{obs}} = 1$.

We prepare a flow with a homogeneous velocity distribution as the
initial configuration, 
\begin{equation}
 v(0,x) = V_i=\text{const.} < 0, \quad c_s(0,x) = 1.
\label{eq:initial}
\end{equation}
This initial configuration of flow has no sonic point and corresponds to
the flat spacetime region before the formation of a black hole. Since our
evolution equations \eqref{eq:num} are of the advection form, it is
enough to set the boundary conditions for $J_+$ at the outlet and $J_{-}$
at the inlet:
\begin{equation}
 J_+(t,-1) = J_+(0,-1) + A_J\cos\left( \om_0 t + \theta \right),
 \quad J_-(t, 1) = J_-(0,1)
\label{eq-ns.bc}
\end{equation}
where $A_J$ is a constant that represents the amplitude of the
perturbation of $J_+$ at $x=-1$, and the values $J_+(0,-1)$ and
$J_-(0,1)$ are determined consistent with Eq.~\eqref{eq:riemann} and
Eq.~\eqref{eq:initial}. These boundary conditions mean that sound
waves of constant amplitude continue to emerge from the outlet toward
the inlet of the nozzle all the time of the numerical simulation.
Here recall that the perturbation $\phi$ of the velocity potential
$\Phi$ plays the role of the scalar field in the ordinary Hawking
radiation. The second term in $J_+(t,-1)$ at the outlet
$A_J\cos\left(\om_0 t + \theta\right)$ causes perturbations of the
velocity potential and results in the classical counterpart of Hawking
radiation at the inlet of the nozzle. Because the amplitude of the
sound wave in one dimensional fluid flow does not decrease, the
amplitude of the input mode of the velocity potential
\eqref{eq:real-input} is given by
\begin{equation}
 A_{\phi}= \left(\frac{c_{s1}+v_1}{2\om_0}\right)A_J,
\end{equation}
where $c_{s1}$ and $v_1$ are respectively the sound velocity and
the fluid velocity at $x=1$, and  the relation of the perturbations
$\phi(t,x=-1) = \delta\Phi(t,-1)
= (c_{s1}+v_1)\int^t dt\, \delta\!J_+(t,-1)/2$
at the outlet is used (see Eqs.~\eqref{eq:riemann} and
\eqref{eq:potential}).

In order to extract the perturbation part $\phi$ from the ``full''
velocity potential $\Phi$ of Eq.~\eqref{eq:potential} which includes
the ``background'' flow of the fluid, we should prepare three input
modes with different initial phases $\theta_1=-\pi/2 , \theta_2=0$
and $\theta_3=\pi/2$. If the amplitude $A_J$ of the perturbation part
of $J_+$ is small enough, then the backreaction effects in the velocity
potentials $\Phi^{\text{out}(\theta_i)}$~($i=1,2,3$) at the observation
point ($x=1$) are negligible and we can subtract the common background
contribution by taking the difference among them. Hence we obtain
the Fourier components of the perturbation (sound wave) at
the observation point as follows
\begin{subequations}
\label{eq:out-signal}
\begin{align}
 \phi_{\omega}^{\text{out}(12)}&\equiv
 \Phi_{\omega}^{\text{out}(\theta_1)}-\Phi_{\omega}^{\text{out}(\theta_2)}=
 \phi_{\omega}^{\text{out}(\theta_1)}-\phi_{\omega}^{\text{out}(\theta_2)},\\
 \phi_{\omega}^{\text{out}(23)}&\equiv
 \Phi_{\omega}^{\text{out}(\theta_2)}-\Phi_{\omega}^{\text{out}(\theta_3)}=
 \phi_{\omega}^{\text{out}(\theta_2)}-\phi_{\omega}^{\text{out}(\theta_3)},
\end{align}
\end{subequations}
where $\Phi_{\om}^{\text{out}(\theta_i)}$ is the Fourier component of
$\Phi^{\text{out}(\theta_i)}$. These Fourier components correspond to
the real valued output signal of Eq.~\eqref{eq:real-output}. Here we note that, 
according to the boundary condition \eqref{eq-ns.bc}, the perturbation
parts of $J_+$ which produce the output signals of
Eqs.~\eqref{eq:out-signal} are given by
\begin{equation}
 \delta J_+^{(12)} = \sqrt{2} A_J \cos\left(\om_0 t + \frac{3\pi}{4} \right)
\quad,\quad
 \delta J_+^{(23)} = \sqrt{2} A_J \cos\left(\om_0 t + \frac{\pi}{4} \right),
\label{eq:input-origin}
\end{equation}
where we used 
$\cos(\om_0 t)-\cos(\om_0 t \pm \pi/2) = \sqrt{2}\cos(\om_0 t \pm \pi/4)$.
Hence we obtain the pure Fourier component
$\phi_{\om}^{\text{out}}$ of the output sound wave by Eq.~\eqref{eq:superpose},
and its power spectrum is given by
\begin{equation}
  |\phi_{\omega}^{\text{out}}|^2=\frac{1}{2}
  \left|\phi_{\omega}^{\text{out}(23)}+i\phi_{\omega}^{\text{out}(12)}\right|^2,
  \label{eq:power}
\end{equation}
where the factor $1/2$ is introduced to eliminate the factor $\sqrt{2}$
which appears in the amplitude $\sqrt{2}A_J$ of the input signals
\eqref{eq:input-origin}. When we carry out a realistic experiment of
the acoustic black hole, the power spectrum \eqref{eq:power} gives
the classical counterpart to Hawking radiation and it has to be compared
with the theoretical form \eqref{eq:planck}. Hence, it is recognized
that we have to perform at least three independent experiments with
three different input phases and observe three independent output signals
to obtain the classical counterpart to Hawking radiation.

With our setting of the fluid flow in the Laval nozzle, we carry
out the numerical simulation by the finite difference method.  We use
the Cubic-Interpolated Pseudoparticle (CIP) method
\cite{YabeT:JCP169:2001} for the interpolation between neighboring
spatial mesh points. This method enables us to treat the shock which
appears after the formation of the sonic point. We note here that
the shock arises in the supersonic region and it does not affect
the subsonic region where we observe the output signal. The
procedure of the numerical simulation to observe the thermal power
spectrum \eqref{eq:planck} of the perturbation of the velocity potential
is as follows:
\begin{description}
  \item[\bf Step 1:] Generate the transonic fluid flow three times 
  with three different initial phase
  values, $\theta_1 = -\pi/2$, $\theta_2 = 0$ and $\theta_3 = \pi/2$.
  Then, using Eq.~\eqref{eq:potential}, obtain the full
  velocity potentials $\Phi^{\text{out}(\theta_i)}$ $(i=1,2,3)$.
  \item[\bf Step 2:] Calculate the quantities
  $\phi^{\text{out}(12)} \equiv
    \Phi^{\text{out}(\theta_1)} - \Phi^{\text{out}(\theta_2)}$
  and
  $\phi^{\text{out}(23)} \equiv
    \Phi^{\text{out}(\theta_2)} - \Phi^{\text{out}(\theta_3)}$.
  Then compute the Fourier components of them,
  $\phi_{\omega}^{\text{out}(12)}$ and $\phi_{\omega}^{\text{out}(23)}$.
\item[\bf Step 3:] 
  Calculate the power spectrum
  \begin{equation}
   P_{\text{obs}}(\om) =
    \frac{1}{2} \left| \phi_{\omega}^{\text{out}(23)}
                     +i\phi_{\omega}^{\text{out}(12)} \right|^2.
   \label{eq:power-exp}
  \end{equation}
  This is the observationally obtained power spectrum for the perturbation
  of the velocity potential.
\item[\bf Step 4:]
  Then compare this power spectrum with the theoretically expected one 
  \begin{equation}
    P_{\text{theory}}(\omega)=\left(\frac{(c_{s1}+v_1)}{2\tilde\omega_0}A_J\right)^2
    \frac{1}{\om_H\,\om}\frac{e^{\om/\om_H}}{e^{\om/\om_H}-1}
  \quad,\quad
   \om_H
   \equiv\frac{\kappa_H}{2\pi}
   =\frac{c_{s0}}{2\pi}\left(\frac{v}{c_s}\right)_{x=x_s}',
   \label{eq:power-th}
  \end{equation}
  and determine the surface gravity $\kappa_H$ of the sonic horizon.
  Here note that the quantity $\tilde\omega_0$ in the amplitude of
  $P_{\text{theory}}$ is the effective input frequency given by the
  following discussion.
\end{description}
We must note here that there arises an extra redshift effect on the
sound wave.  In our numerical simulation, the initial fluid flow has
homogeneous velocity distribution.  Then the system starts to evolve
dynamically and settles down to the stationary transonic fluid flow
finally. Therefore, the observed frequency shifts due to the evolution
of the ``background'' flow. According to the null coordinate
\eqref{eq:null1}, the outgoing wave with frequency $\omega_0$ is
\begin{equation}
  \phi(t,x=1)\propto\exp\left(-i\omega_0\left(t-\int_{-1}^{1}
\frac{dx}{c_s+v}\right)\right),
\end{equation}
and the effective frequency of the wave at $x=1$ is given by
\begin{equation}
  \label{eq:eff-omega}
  \tilde\omega_0=\omega_0\left(1
    -\frac{\partial}{\partial t}\int_{-1}^{1}\frac{dx}{c_s+v}\right).
\end{equation}
The second term in Eq.~\eqref{eq:eff-omega} gives a negative
contribution in our non-stationary setting of numerical experiment
and the observed frequency $\tilde\omega_0$ at $x=1$ becomes smaller
than $\omega_0$. The theoretically expected form of the power spectrum
in our setting should be given by replacing $\omega_0$ to
$\tilde\omega_0$ in the formula \eqref{eq:planck}.

\section{\label{sec:result}Numerical results}

We use the parameters $\gamma = 7/5$, $A_J = 1.0 \times 10^{-6}$ and
$\om_0 = 100$. The number of spatial mesh points are 10001 and the
size of one mesh becomes $\Delta x = 2/10001$. The size of one
temporal step is set $\Delta t = \Delta x/10$, and we calculate 900000
steps in time. The numerical simulation runs in the temporal range $t
= 0 \sim 18$.  The spatial scale is normalized by $L/2$, where $L$ is
the length of the nozzle. The temporal scale is normalized by $L/2
c_s(t=0)$, where we set $c_s(t=0) = 1$ as denoted by the initial
condition \eqref{eq:initial}.

\subsection{no black hole  case}
\begin{figure}[H]
  \centering
  \includegraphics[width=0.45\linewidth,clip]{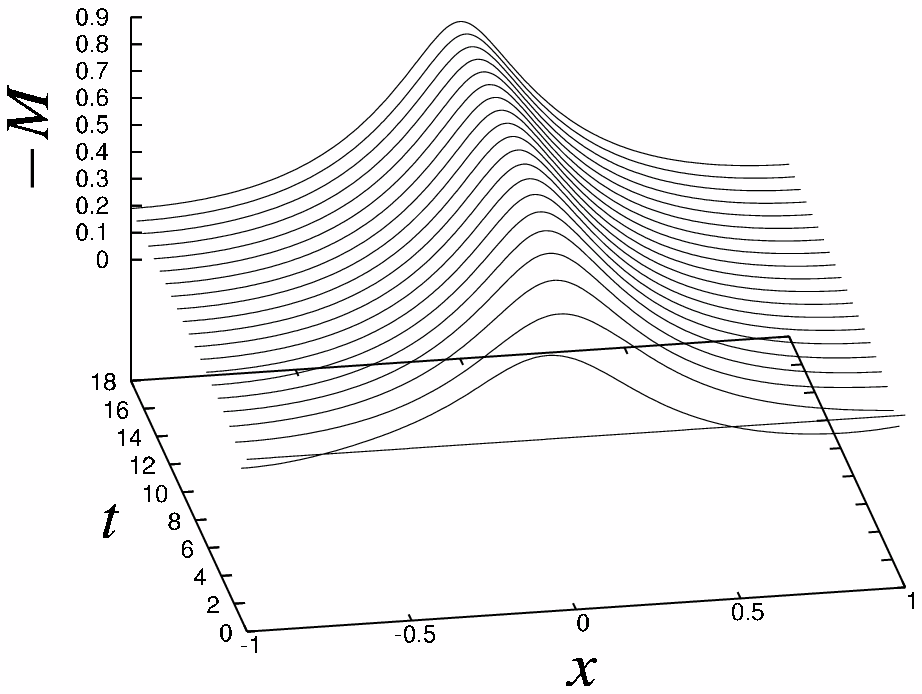}\hspace{1cm}
  \includegraphics[width=0.45\linewidth,clip]{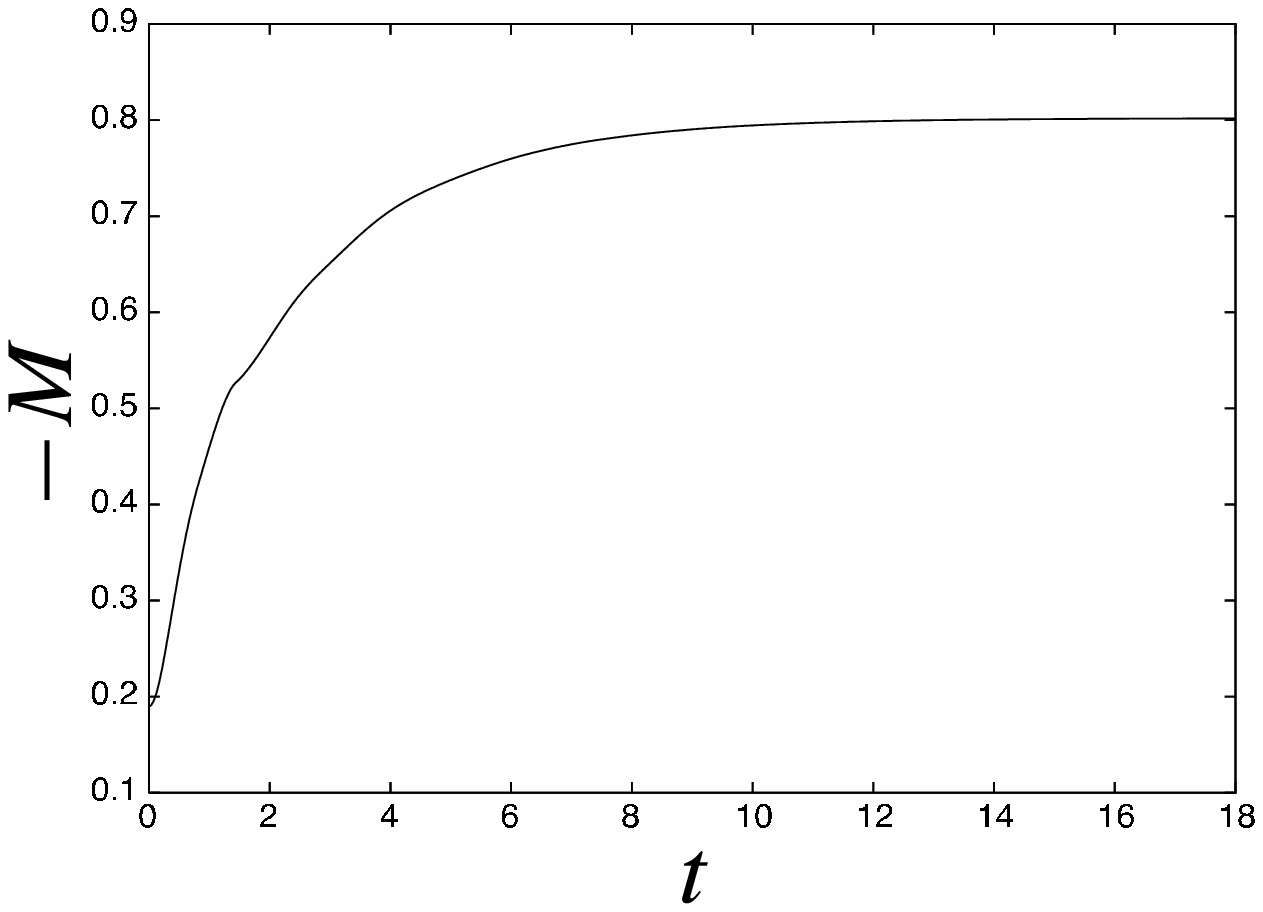}
  \caption{The left panel shows the temporal evolution of the spatial
    distribution of the Mach number. The right panel shows the Mach
    number at $x=0$.}
  \label{fig:mach1}
\end{figure}

For the initial fluid velocity $V_i=-0.19$, we do not observe the
formation of an acoustic black hole. Figure~\ref{fig:mach1} shows the
evolution of the Mach number.  After $t\sim 10$, the Mach number at
$x=0$ reaches a constant value and a stationary flow is realized.
\begin{figure}[H]
  \centering
  \includegraphics[width=0.5\linewidth,clip]{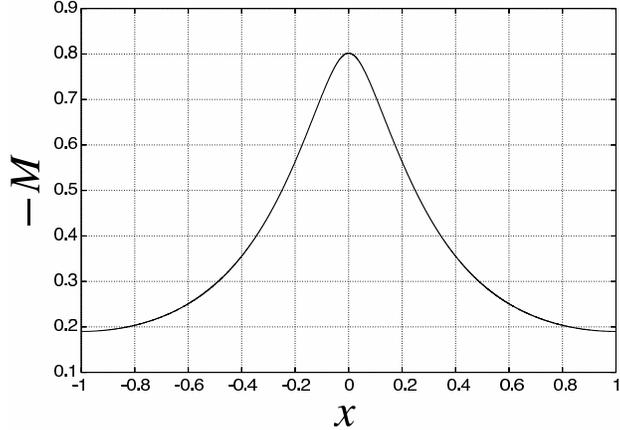}
  \caption{Spatial distribution of the Mach number at $t=18$.}
  \label{fig:final-mach1}
\end{figure}

\begin{figure}[H]
  \centering
  \includegraphics[width=0.45\linewidth,clip]{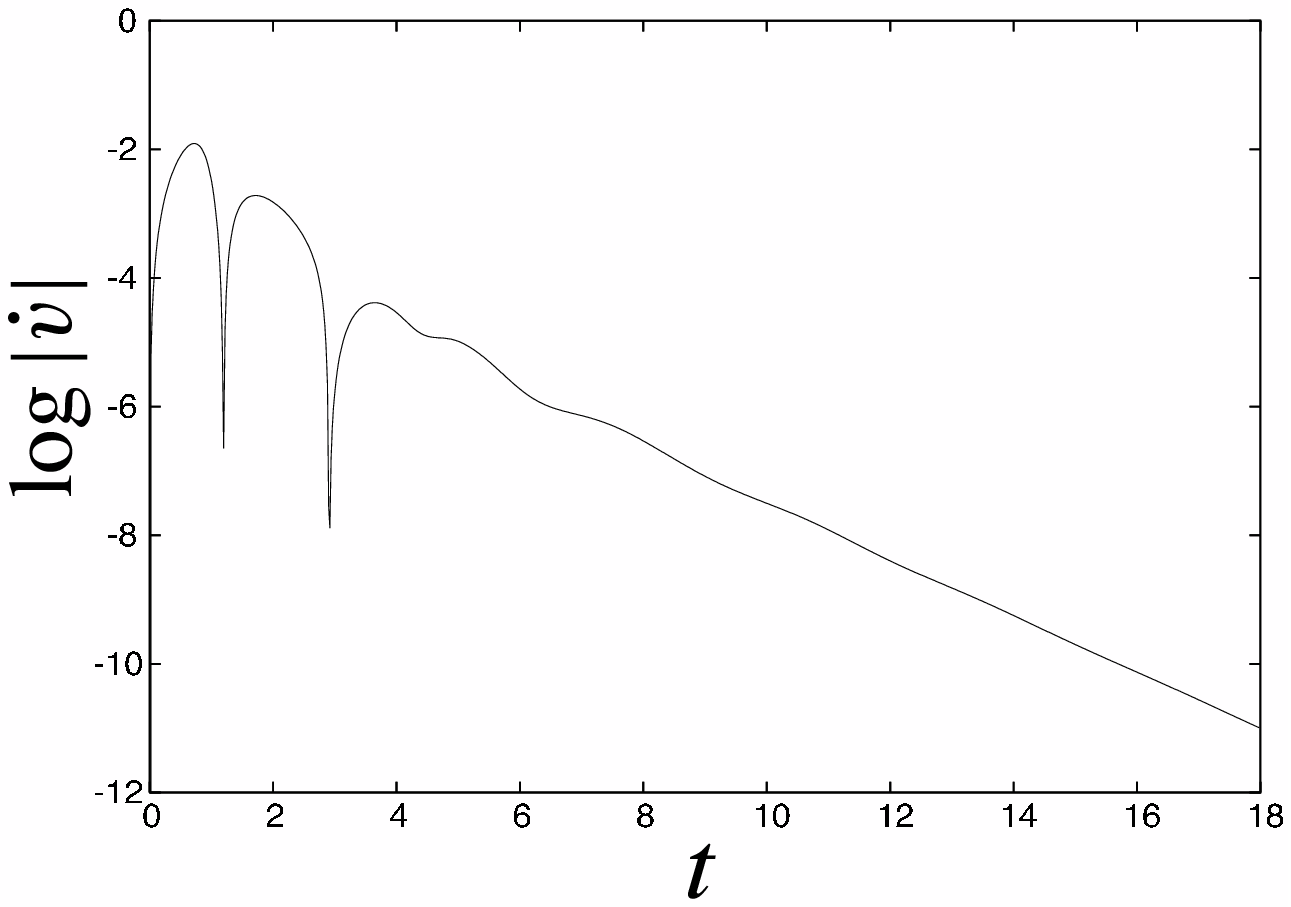}
 \includegraphics[width=0.45\linewidth,clip]{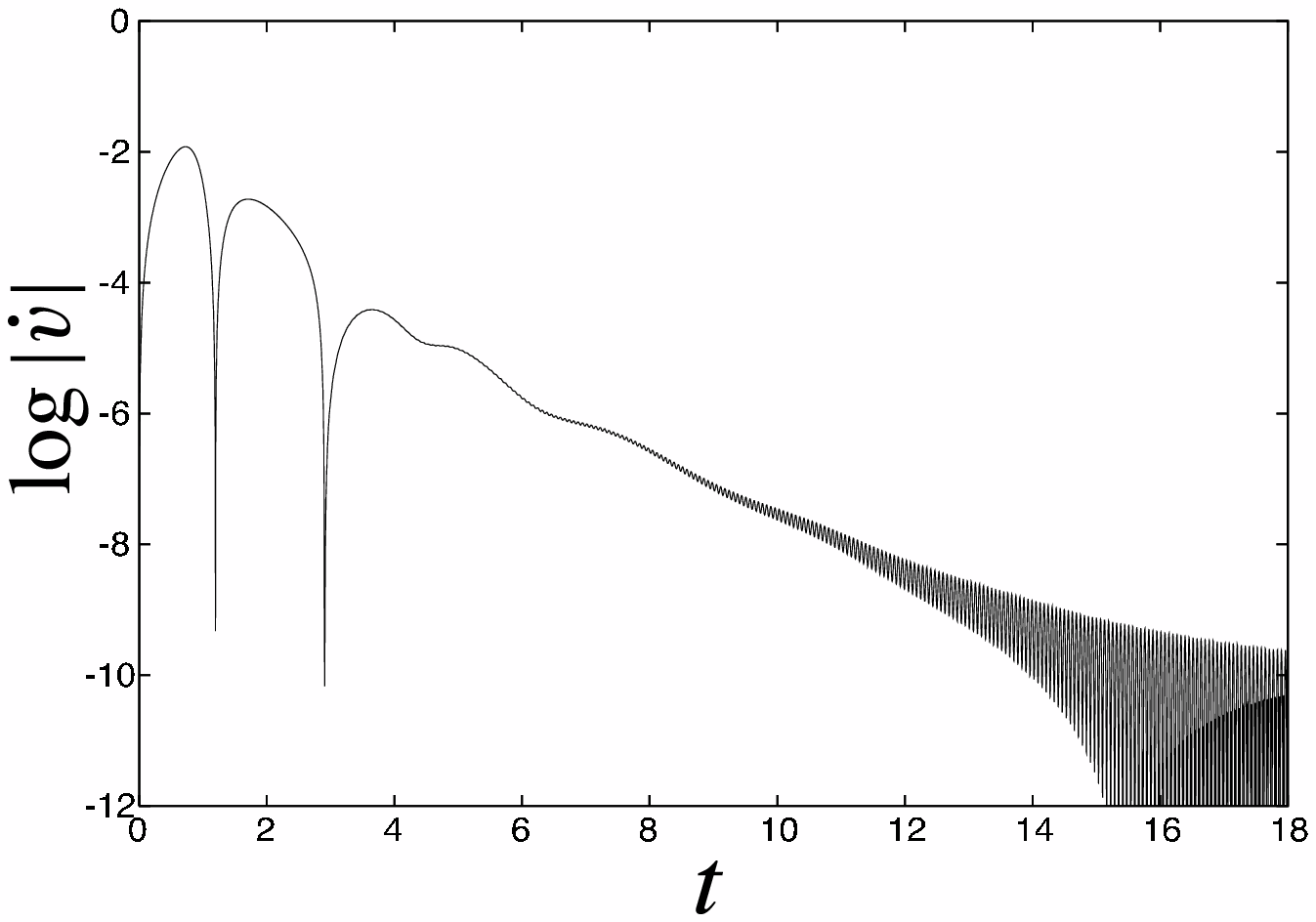}
  \caption{The temporal evolution  of $\dot v$ at $x=1$. The left
    panel is $A_J=0$ case and the right panel is $A_J\neq 0$ case.}
  \label{fig:dv1}
\end{figure}

\noindent
The spatial distribution of the fluid velocity at $t=18$
(FIG.~\ref{fig:final-mach1}) coincides with the stationary solution
\eqref{eq:stationary} with $M_{\text{in}}=-0.19$ and no sonic point
appears. Figure \ref{fig:dv1} shows the time derivative of the observed
velocity at $x=1$ which should disappear when the fluid flow becomes
stationary. Until $t\sim 4$, the initial burst mode appears, which is
peculiar to our initial and boundary conditions. Then, the output
signal decays exponentially. For $A_J\neq 0$ (with input perturbation),
the output signal oscillates with a constant amplitude after the
``background'' flow settles down to the stationary flow.

\begin{figure}[H]
  \centering
\includegraphics[width=0.65\linewidth,clip]{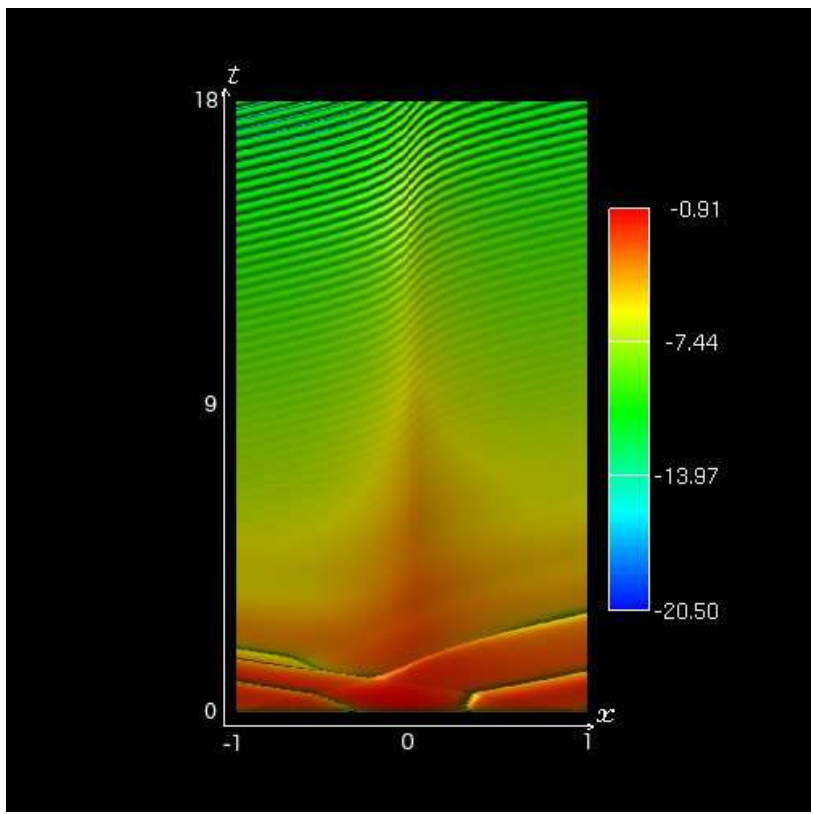}
\caption{Spacetime distribution of $\dot v$ for
  $V_i=-0.19$ with $A_J\neq 0, \omega_0=20$. Colors are assigned
  according to the value of $\log|\dot v|$.}
  \label{fig:diagram1}
\end{figure}

\begin{figure}[H]
  \centering
\includegraphics[width=0.65\linewidth,clip]{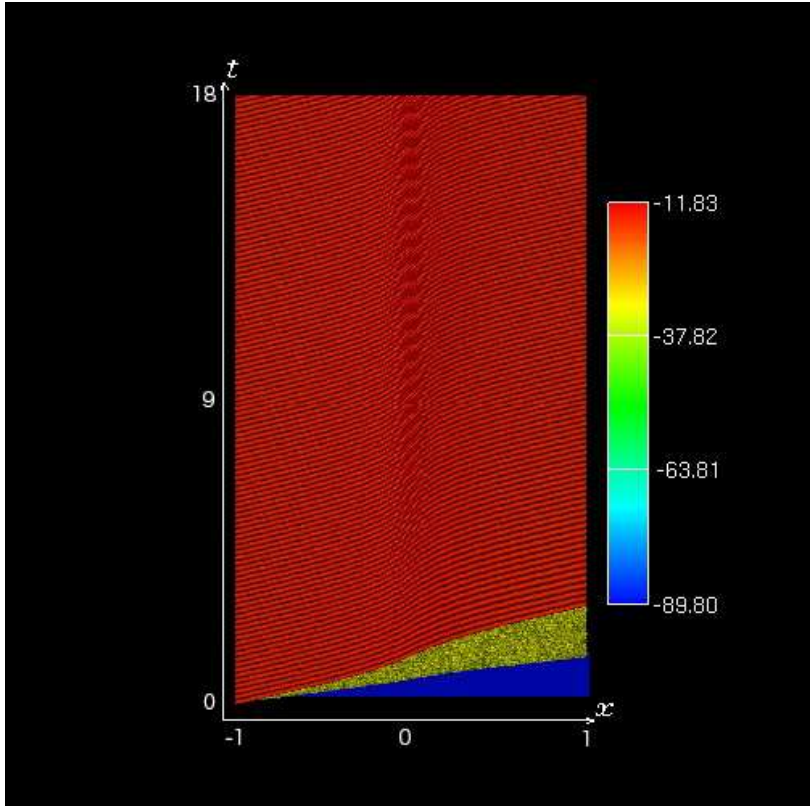}
\caption{Spacetime distribution of the velocity perturbation $\delta
  v\equiv v^{(2)}-v^{(1)}$ for $V_i=-0.19$ with $A_J\neq 0, \omega_0=20$.
  Colors are assigned according to the value of $\log|\delta v|$.}
  \label{fig:diagram11}
\end{figure}

Figures~\ref{fig:diagram1} and \ref{fig:diagram11}  show the spacetime
distribution of the velocity perturbation for $A_J\neq 0$ with
$\omega_0=20$.  \footnote{For FIG.~\ref{fig:diagram1},
  \ref{fig:diagram11},\ref{fig:diagram2} and \ref{fig:diagram22}, we
  used  the smaller value of the perturbation frequencty $\omega_0=20$
  to visualize the propagation of the wave in the spacetime diagram. }
In FIG.~\ref{fig:diagram11}, to subtract the ``background'' flow, we
have defined $\delta v=v^{(2)}-v^{(1)}$ where $v^{(i)}$ denotes the
fluid velocity with the input phase $\theta_i$. We can see that the
sound wave propagates from $x=-1$ to $x=1$. Around the throat $x=0$,
the fluid velocity has larger value compared to the other region and
the phase velocity of the outgoing wave becomes smaller. Thus the
slope of the constant phase line becomes larger  around the throat.

\begin{figure}[H]
  \centering
  \includegraphics[width=0.45\linewidth,clip]{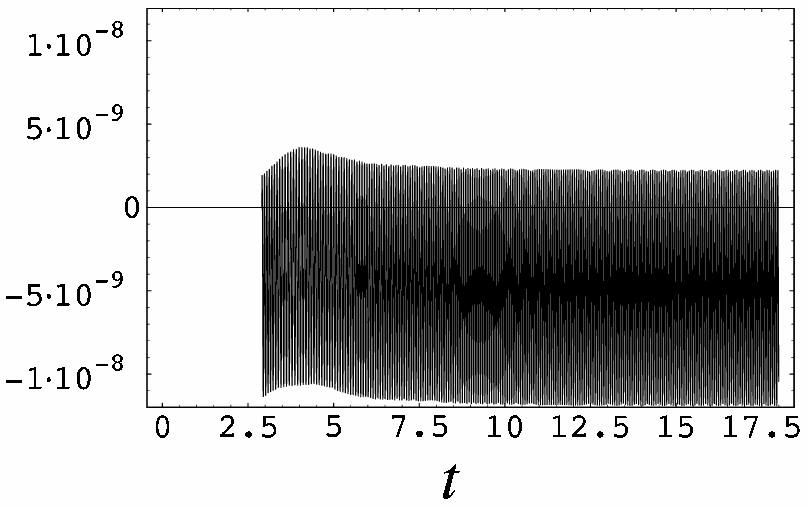}%
  \includegraphics[width=0.45\linewidth,clip]{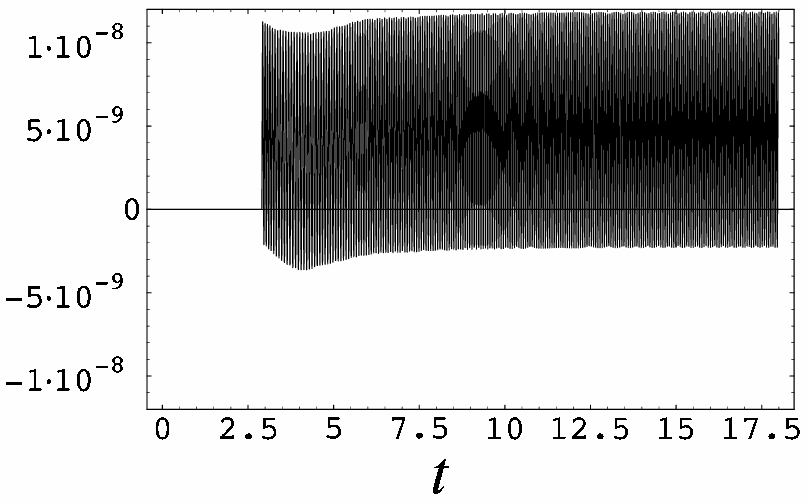}
  \caption{The temporal evolution of the velocity potential $\phi^{\text{out}(12)}$ and
    $\phi^{\text{out}(23)}$.}
  \label{fig:phi1}
\end{figure}

Figure~\ref{fig:phi1} shows the perturbation of the velocity potential at
$x=1$. We set $t_0=3$ in Eq.~\eqref{eq:potential} to obtain the
velocity potential. The observed perturbation of the velocity
potential oscillates with a constant amplitude.
\begin{figure}[H]
  \centering
  \includegraphics[width=0.45\linewidth,clip]{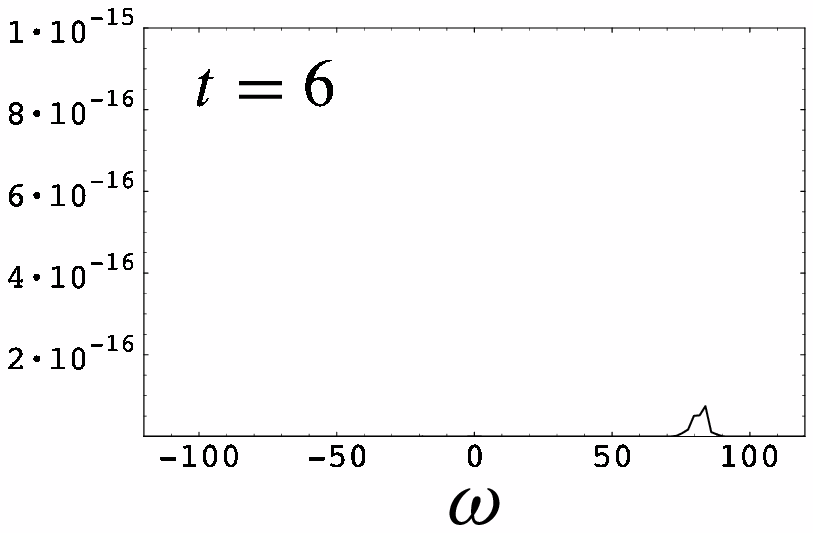}%
\includegraphics[width=0.45\linewidth,clip]{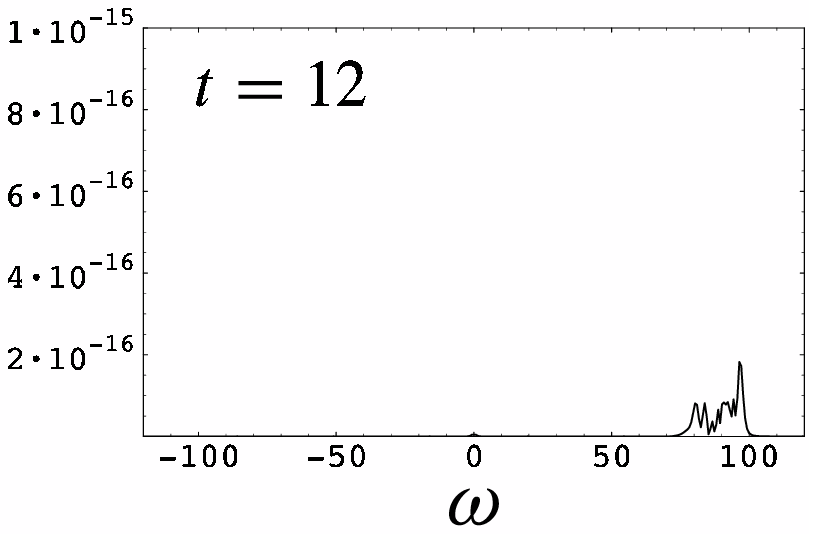}
\includegraphics[width=0.45\linewidth,clip]{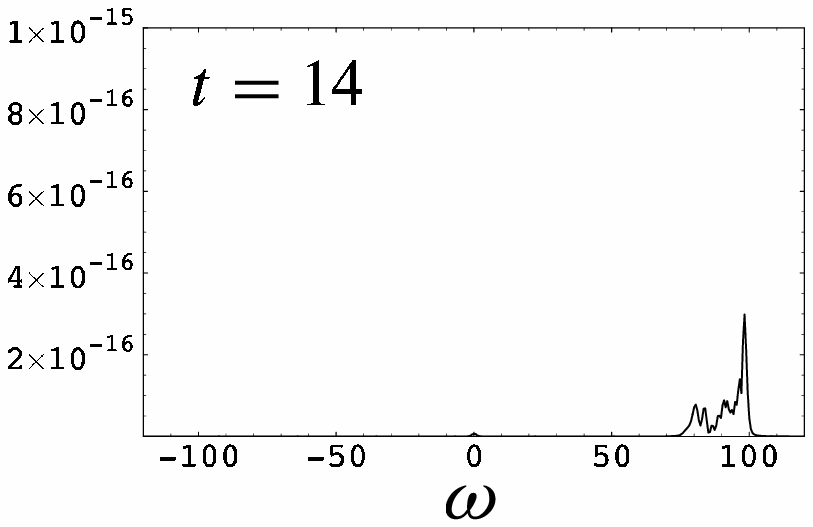}%
\includegraphics[width=0.45\linewidth,clip]{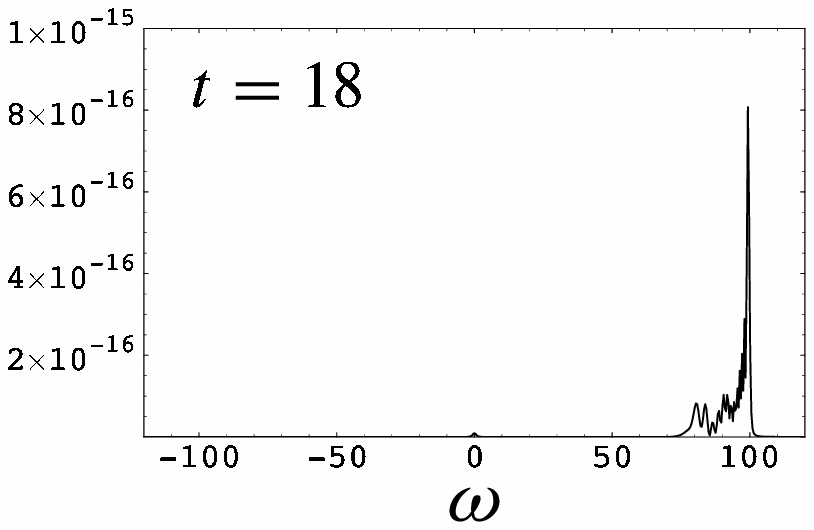}
  \caption{The temporal evolution of the power spectrum $P_{\text{obs}}(\omega)$ for the %
     velocity potential.}
   \label{fig:power1}
\end{figure}
The power spectrum $P_{\text{obs}}(\omega)$ obtained from
the perturbations are shown in FIG.~\ref{fig:power1}. The frequency
of the observed signal evolves in time. This is due to
the non-stationarity of the background flow as denoted by
Eq.~\eqref{eq:eff-omega}. Until the flow settles down to the stationary
flow which is determined by the given boundary condition, the background
flow evolves in time and causes the shift of the frequency of
the observed perturbation. After a sufficiently long time has passed and
the fluid flow becomes stationary, the observed frequency coincides with
the input frequency $\omega_0=100$ as expected by Eq.~\eqref{eq:eff-omega}.

\subsection{ black hole formation case}
\begin{figure}[H]
  \centering
  \includegraphics[width=0.45\linewidth,clip]{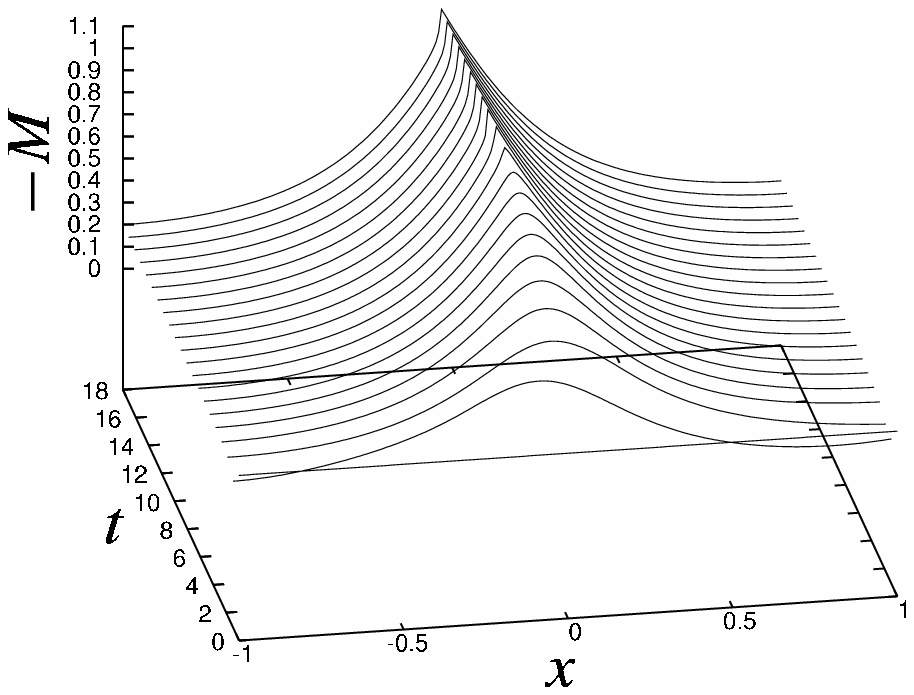}\hspace{0.5cm}
  \includegraphics[width=0.45\linewidth,clip]{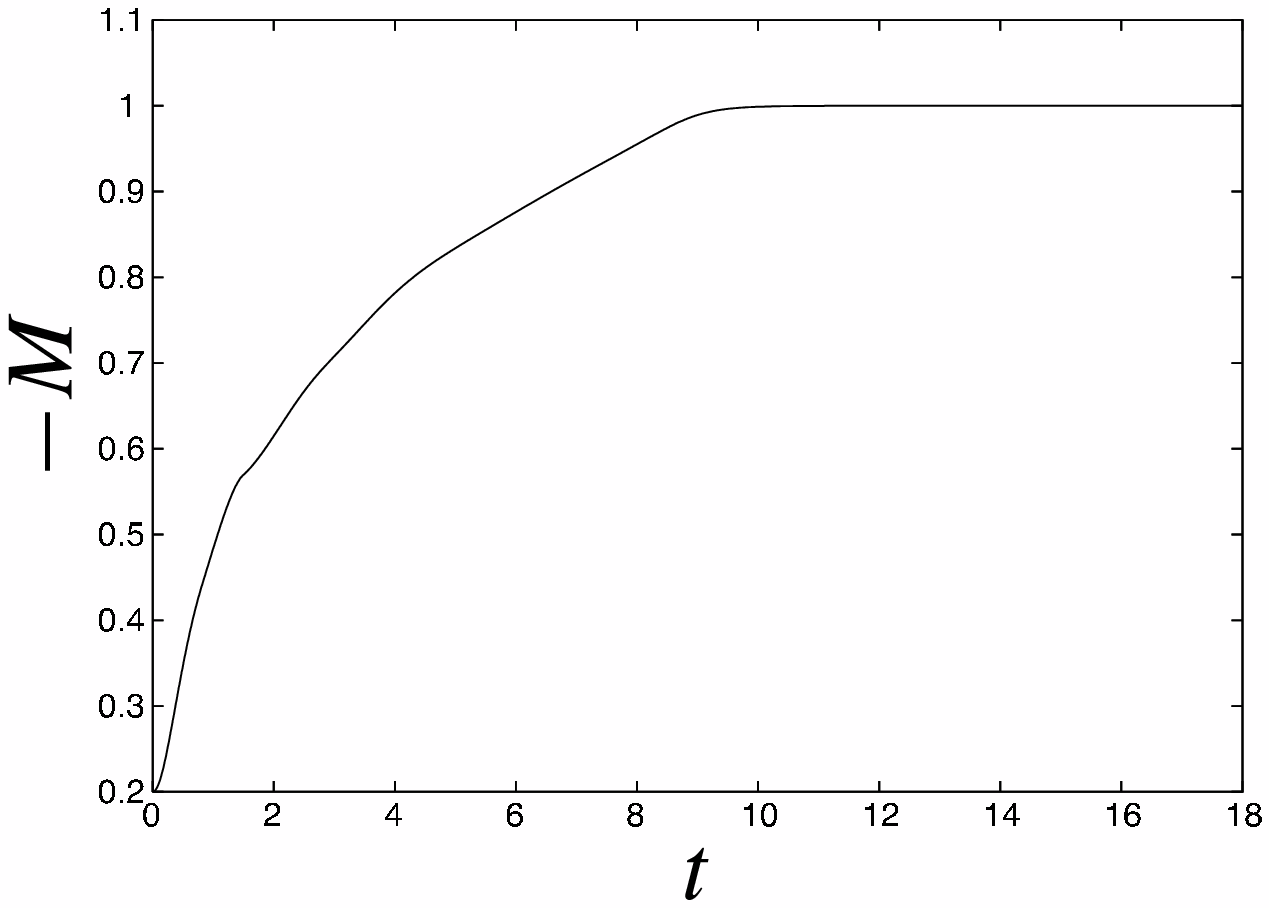}
  \caption{The left panel shows the temporal evolution of the spatial
    distribution of the Mach number. The right panel shows the Mach
    number at $x=0$.}
  \label{fig:mach2}
\end{figure}

For the initial fluid velocity $V_i=-0.2$, we observe the formation
of the acoustic black hole. Figure~\ref{fig:mach2} shows the evolution of
the Mach number. 
\begin{figure}[H]
  \centering
  \includegraphics[width=0.5\linewidth,clip]{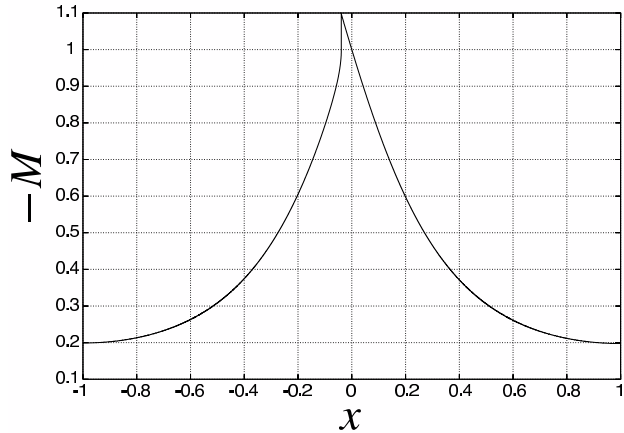}
  \caption{Spatial distribution of the Mach number at $t=18$.}
  \label{fig:final-mach2}
\end{figure}
\noindent
At $t\sim10$, the Mach number at the throat $x=0$ reaches unity and
the sonic point appears (formation of the acoustic black hole). At the
same time, the discontinuity of the derivative of the Mach number appears
in the supersonic region $x<0$ as shown in FIG.~\ref{fig:final-mach2}.
This discontinuity is due to the shock formed in the transonic flow.
The formation of the shock is peculiar to our boundary condition which 
determines the Riemann invariants $J_\pm$ at the inlet and the outlet
of the nozzle. However, since the shock occurs in the supersonic region
after the formation of the sonic point, the effect of the shock never
propagate into the subsonic region after the formation of the sonic
point. Hence the shock never affect the sonic analogue of Hawking
radiation, and we do not have to pay attention to the effect of the
shock formation. 

\begin{figure}[H]
  \centering
  \includegraphics[width=0.45\linewidth,clip]{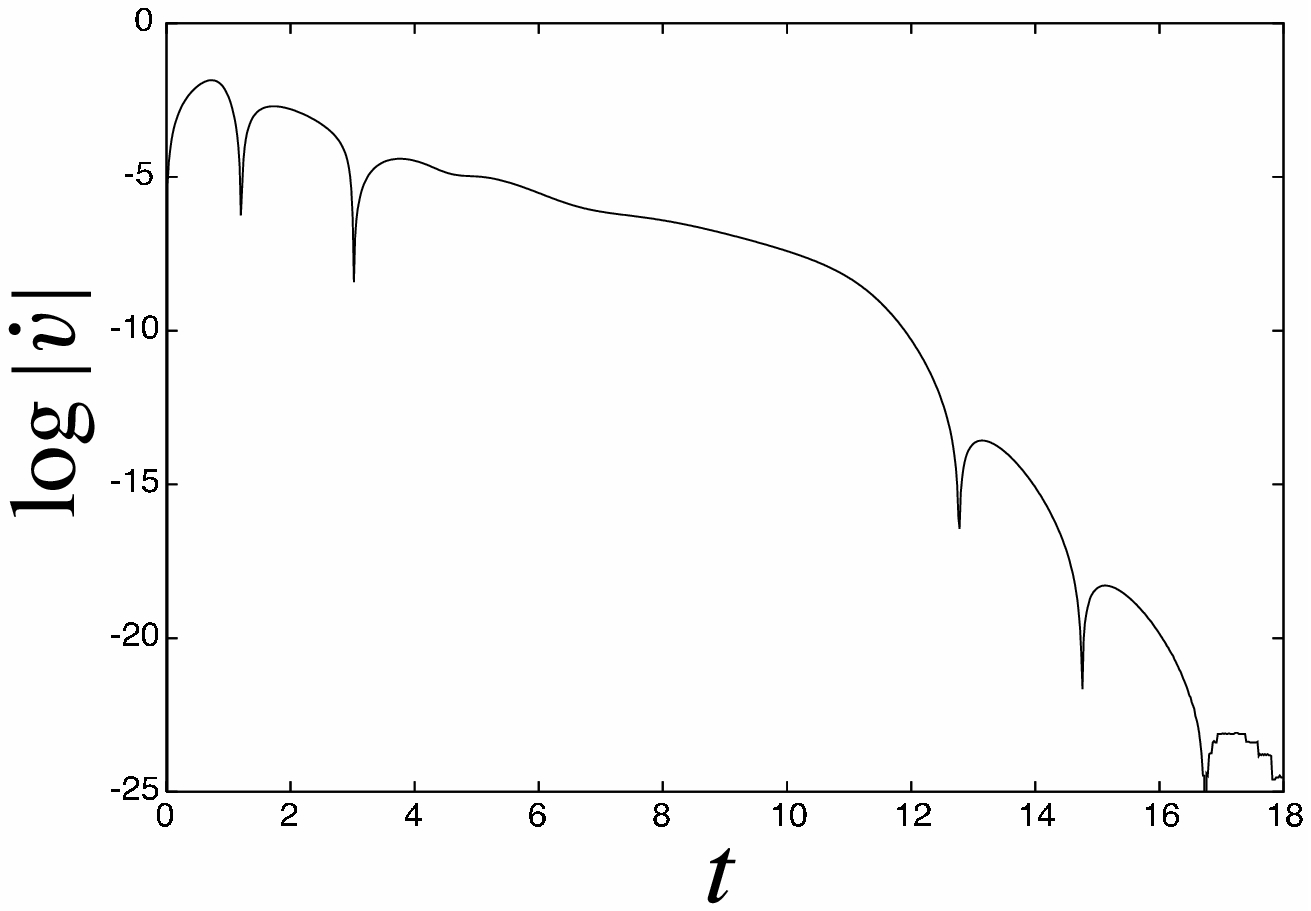}
  \includegraphics[width=0.45\linewidth,clip]{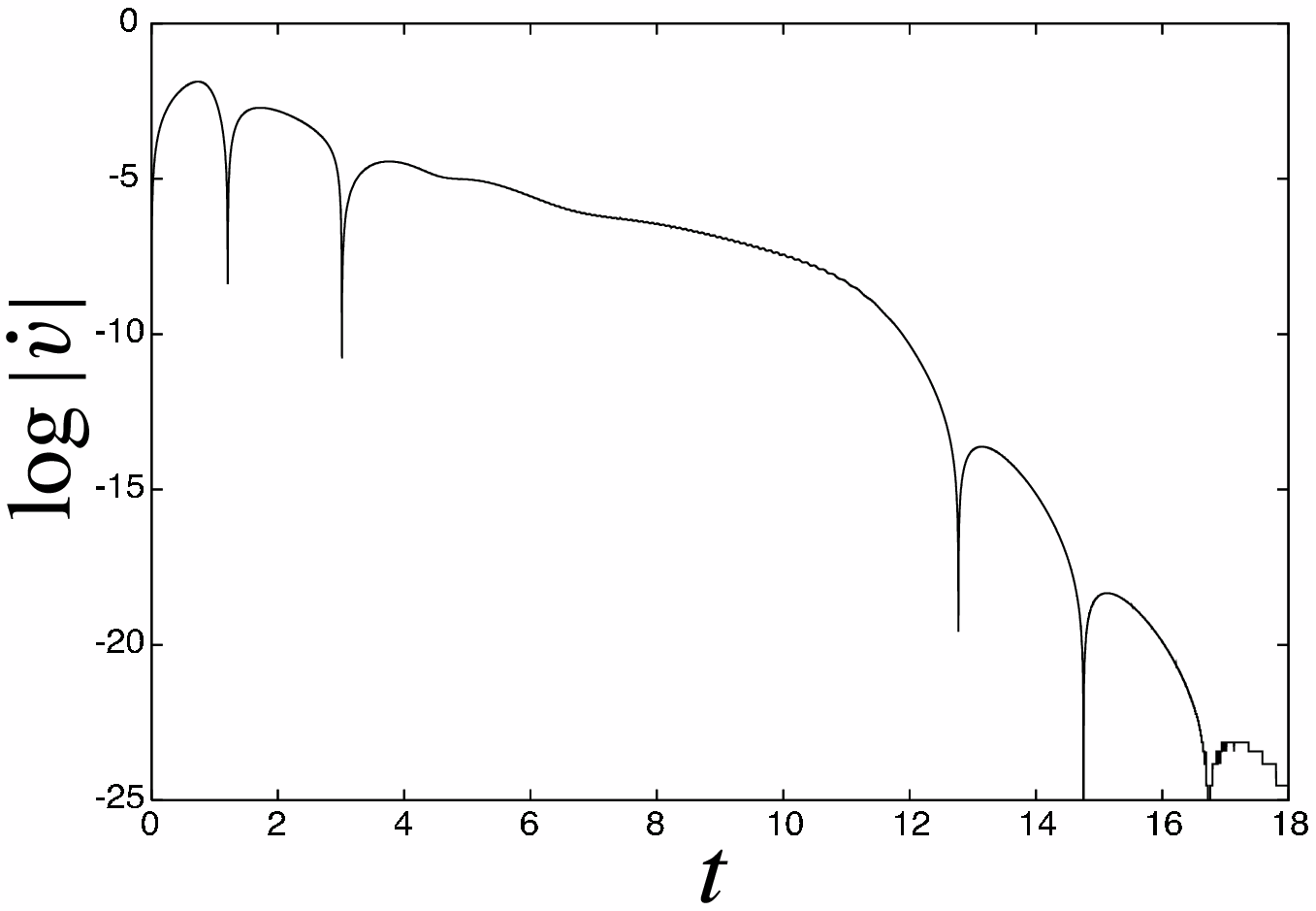}
   \caption{The temporal evolution  of $\dot v$ at $x=1$. The left
     panel is $A_J=0$ case and the right panel is $A_J\neq 0$ case. As
   the amplitude of the input perturbation is far smaller than the change
   of the background, we can not see the oscillation of the
   perturbation in the figure (right panel).}
   \label{fig:dv2}
\end{figure}

Figure~\ref{fig:dv2} is the time derivative of the fluid velocity at
$x=1$. After the burst mode $t\sim 4$, we can observe the quasi-normal
oscillation \cite{FrolovYP:1998} of the acoustic black hole with
a period $\sim 4$.
\noindent
The value of this period can be estimated as follows. In our
simulation, the value of $J_{-}$ at the inlet of the nozzle is fixed
to be constant by the boundary condition. Hence after the formation
of the sonic horizon, the boundary condition for the ingoing
perturbation $\del J_{-}$ becomes
\begin{equation}
  \delta J_{-}=\text{free} : \text{ at the horizon},\quad \delta J_{-}=0 :
  \text{at the inlet of the nozzle}
\end{equation}
For the outgoing perturbation $\del J_{+}$,
\begin{equation}
  \delta J_{+}=0 : \text{ at the horizon},\quad \delta
  J_{+}=\text{free} : \text{ at the inlet of the nozzle}
\end{equation}
By these boundary condition, a quarter of the wavelength of the
quasi-normal mode is equal to a half length of the Laval nozzle and the
wavelength of the quasi-normal mode becomes 4 in the present case.
Thus, we obtain the period 4 of the quasi-normal mode by assuming
$c_s=1$.

\begin{figure}[H]
  \centering
  \includegraphics[width=0.65\linewidth,clip]{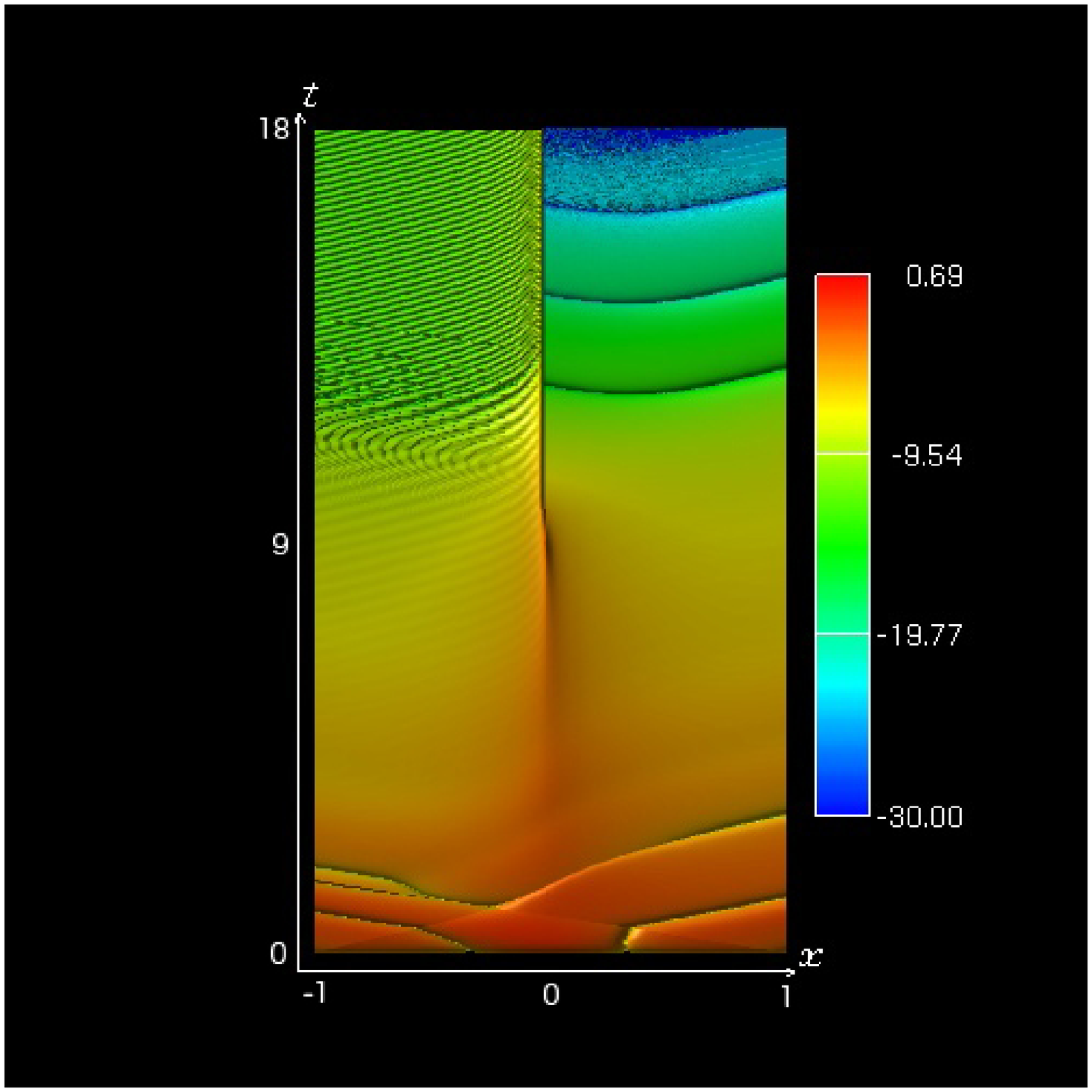}
  \caption{Spacetime distribution of $\dot v$ for $V_i=-0.2$ with
    $A_J\neq0, \omega_0=20$. Colors are assigned according to the value
    of $\log|\dot v|$.}
  \label{fig:diagram2}
\end{figure}

\begin{figure}[H]
  \centering
  \includegraphics[width=0.65\linewidth,clip]{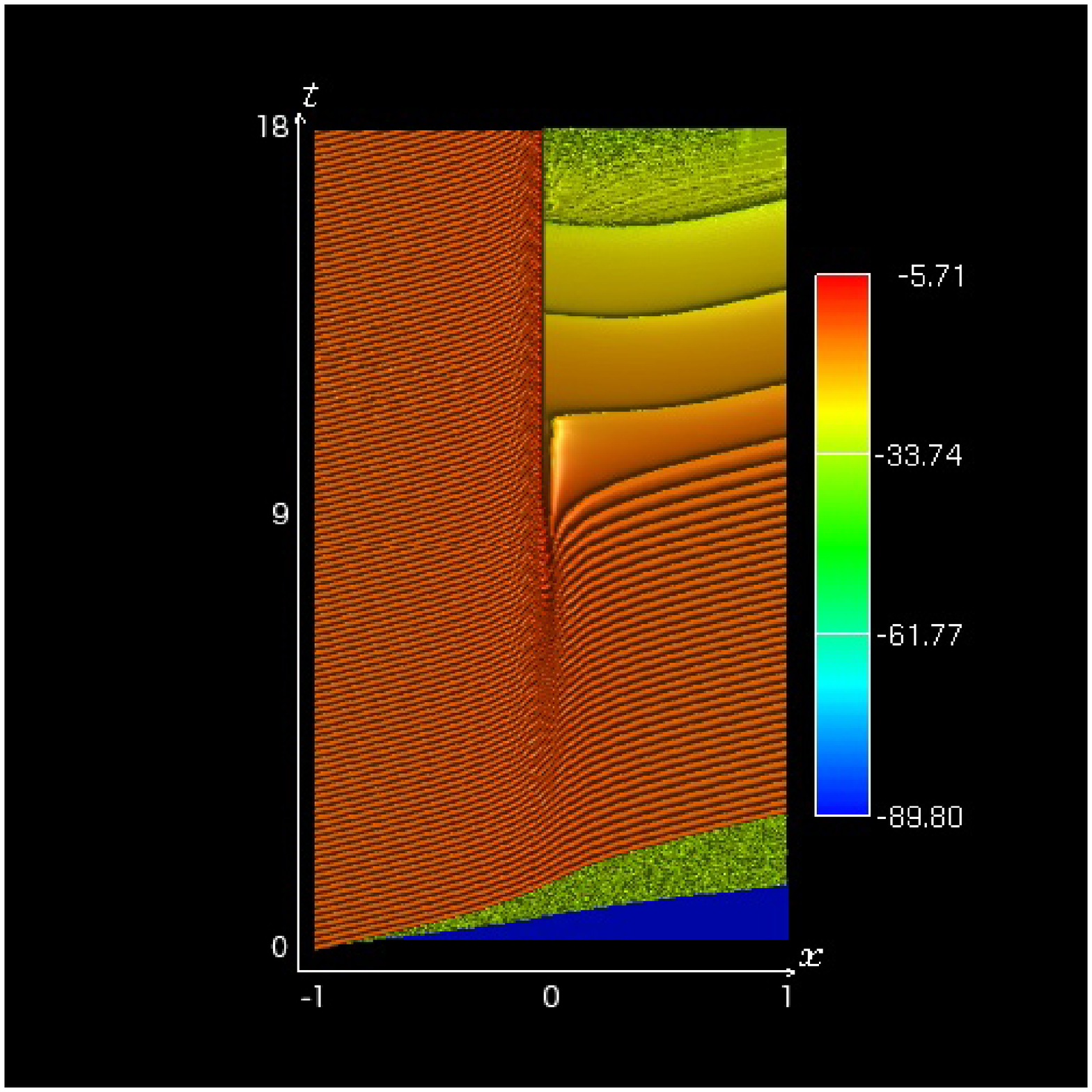}
  \caption{Spacetime distribution of the velocity perturbation $\delta
    v=v^{(2)}-v^{(1)}$ for $V_i=-0.2$ with
    $A_J\neq0, \omega_0=20$. Colors are assigned according to the value
    of $\log|\delta v|$.}
  \label{fig:diagram22}
\end{figure}

The spacetime distribution of the time derivative of the fluid velocity
and the perturbation is shown in FIG.~\ref{fig:diagram2} and
FIG.~\ref{fig:diagram22}, respectively. 

\begin{figure}[H]
  \centering
  \includegraphics[width=0.45\linewidth,clip]{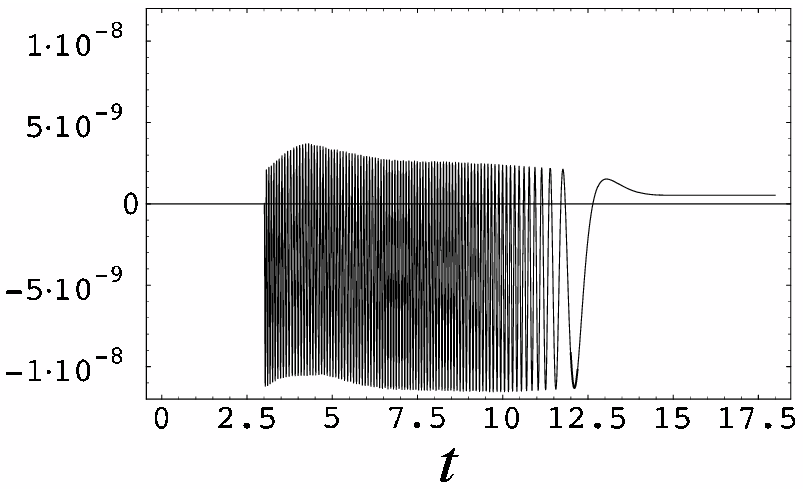}
  \includegraphics[width=0.45\linewidth,clip]{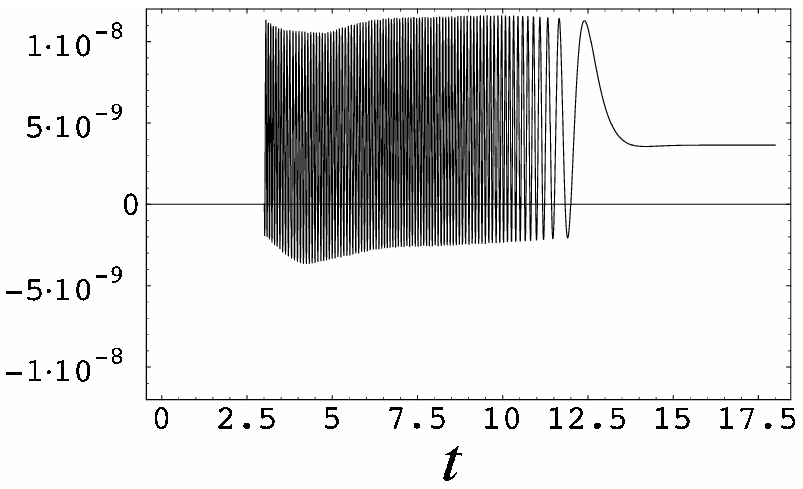}
  \caption{The temporal evolution of the velocity potential $\phi^{\text{out}(12)}$ and
    $\phi^{\text{out}(32)}$.}
  \label{fig:phi2}
\end{figure}

The time evolution of the observed perturbation of the velocity
potential is shown in FIG.~\ref{fig:phi2}. After the formation of
the sonic horizon $t\sim10$, the frequency of the observed perturbation
gradually decreases to zero.

\begin{figure}[H]
 \centering
\includegraphics[width=0.45\linewidth,clip]{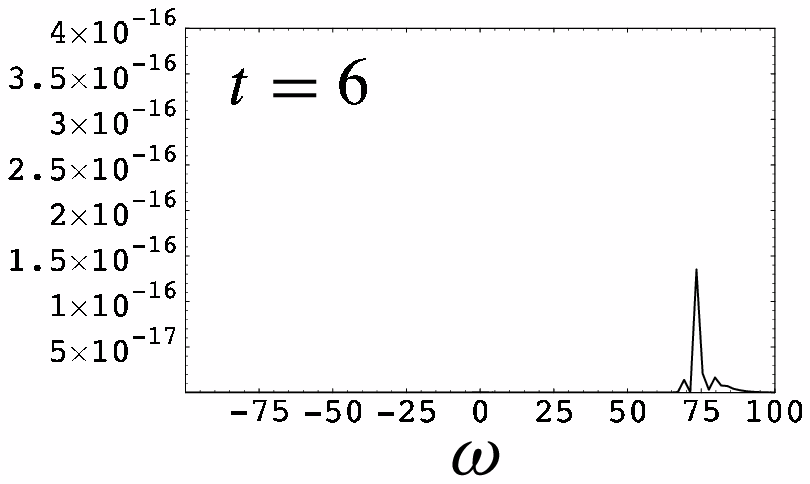}%
\includegraphics[width=0.45\linewidth,clip]{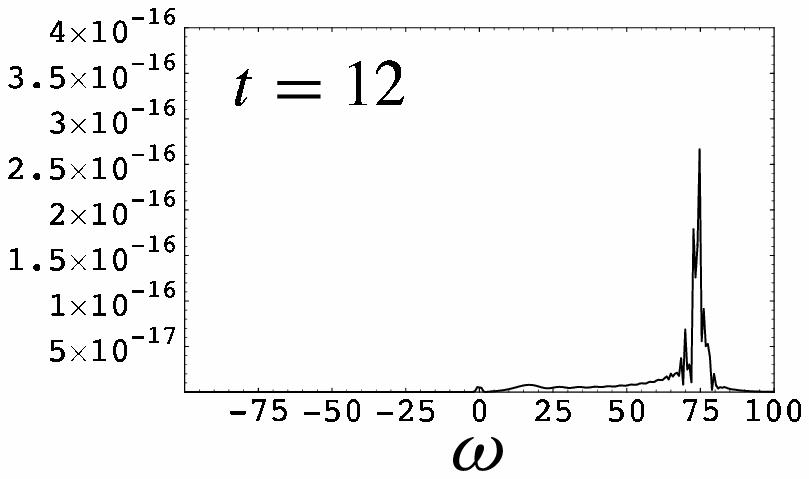}
\includegraphics[width=0.45\linewidth,clip]{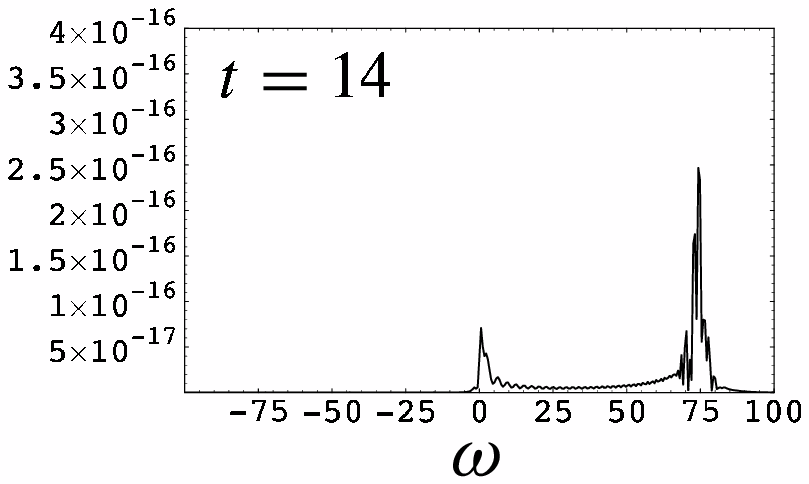}%
\includegraphics[width=0.45\linewidth,clip]{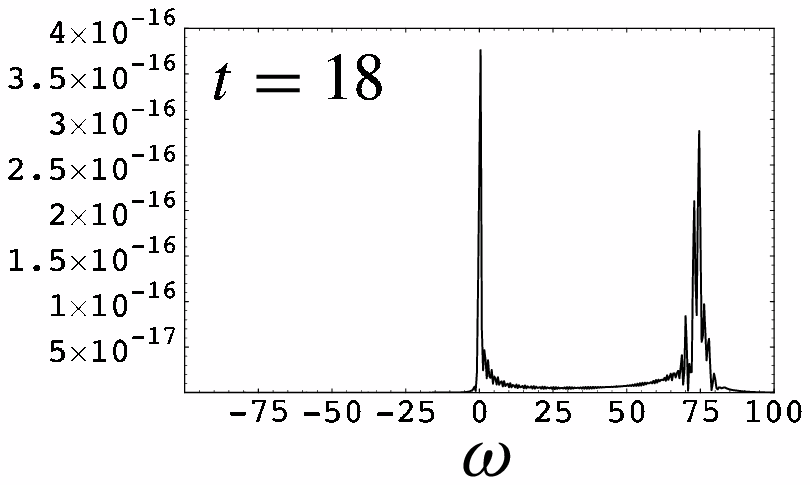}
  \caption{The temporal evolution of the power spectrum $P_{\text{obs}}(\omega)$ for the %
     output signal.}
   \label{fig:power2}
\end{figure}

\begin{figure}[H]
  \centering
\includegraphics[width=0.45\linewidth,clip]{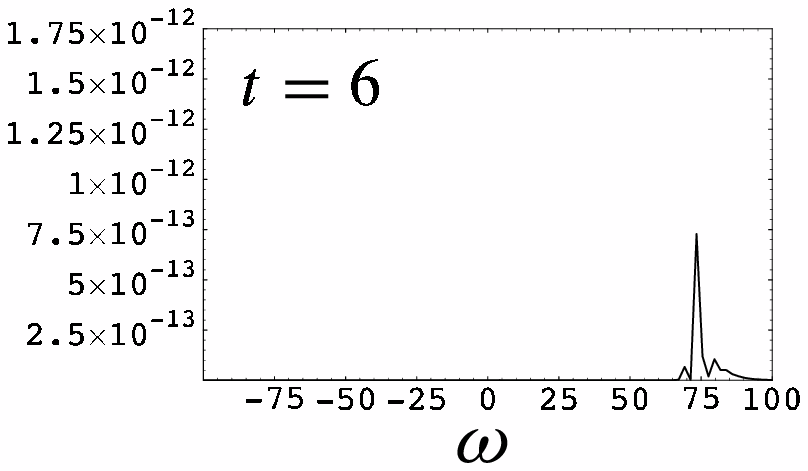}%
\includegraphics[width=0.45\linewidth,clip]{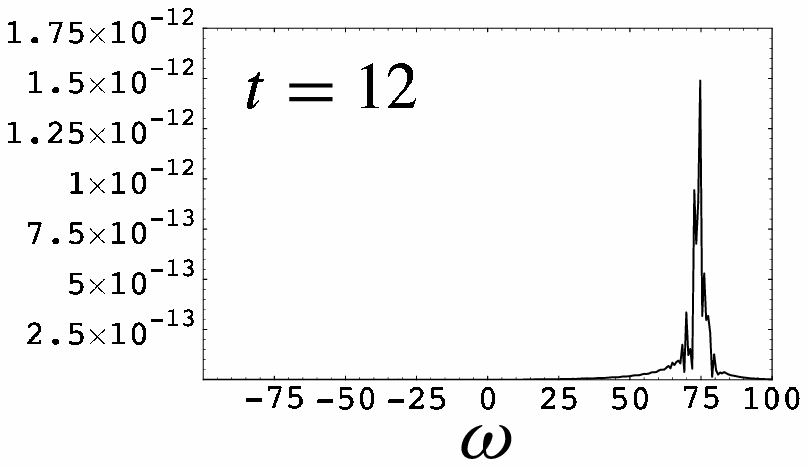}
\includegraphics[width=0.45\linewidth,clip]{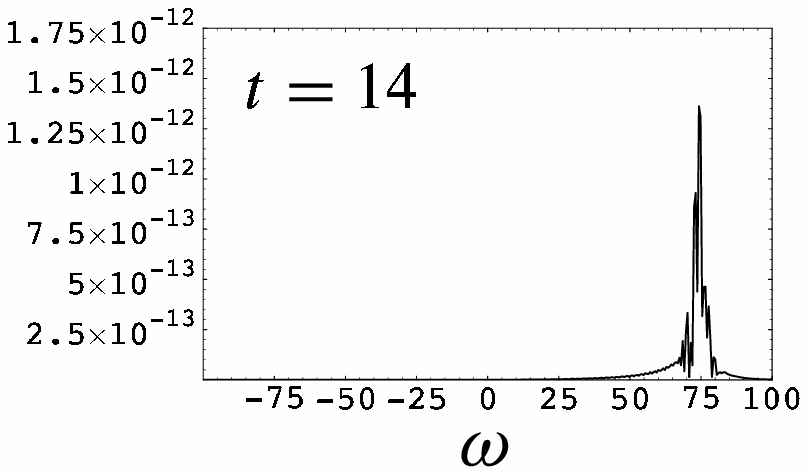}%
\includegraphics[width=0.45\linewidth,clip]{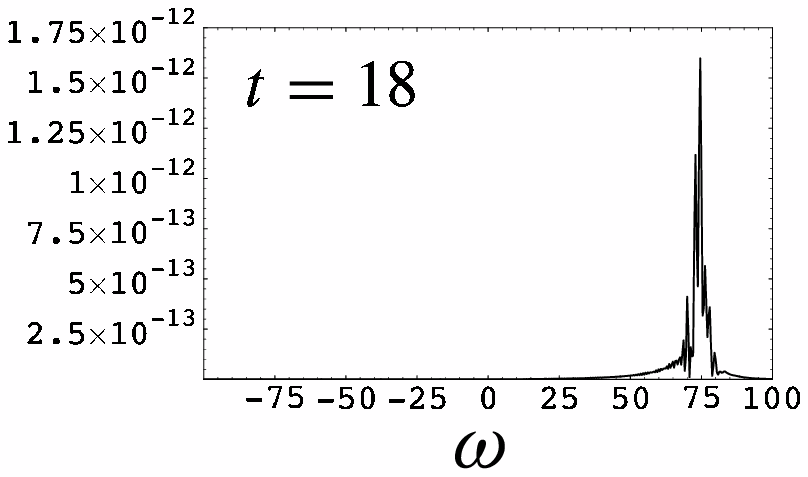}
  \caption{The temporal evolution of the power spectrum $\omega^2P_{\text{obs}}(\omega)$ for the %
     output signal.}
   \label{fig:power3}
\end{figure}

\begin{figure}[H]
  \centering
  \includegraphics[width=0.45\linewidth,clip]{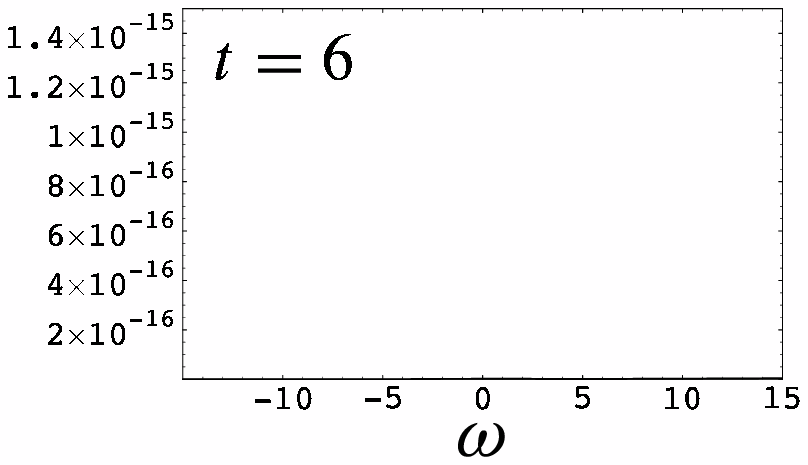}%
  \includegraphics[width=0.45\linewidth,clip]{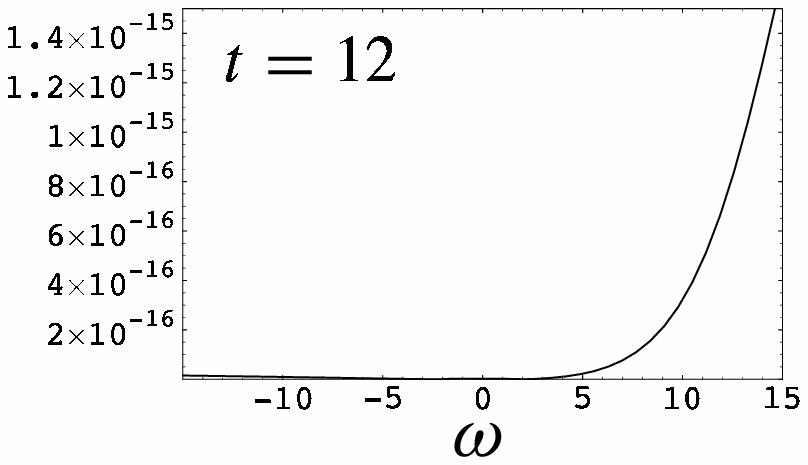}
  \includegraphics[width=0.45\linewidth,clip]{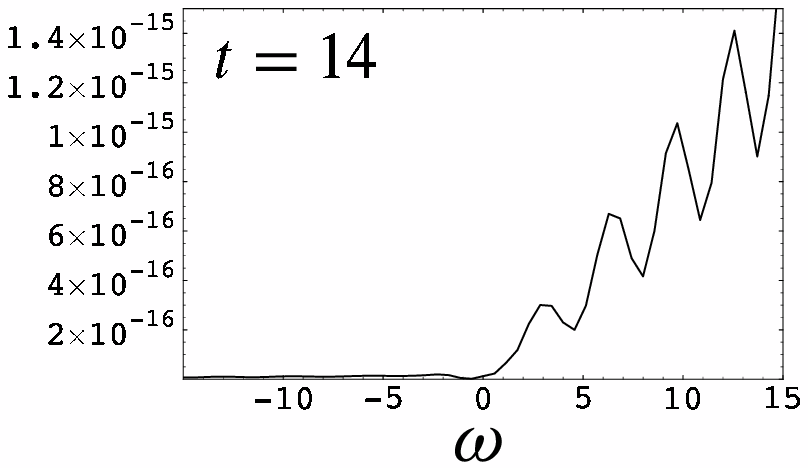}%
  \includegraphics[width=0.45\linewidth,clip]{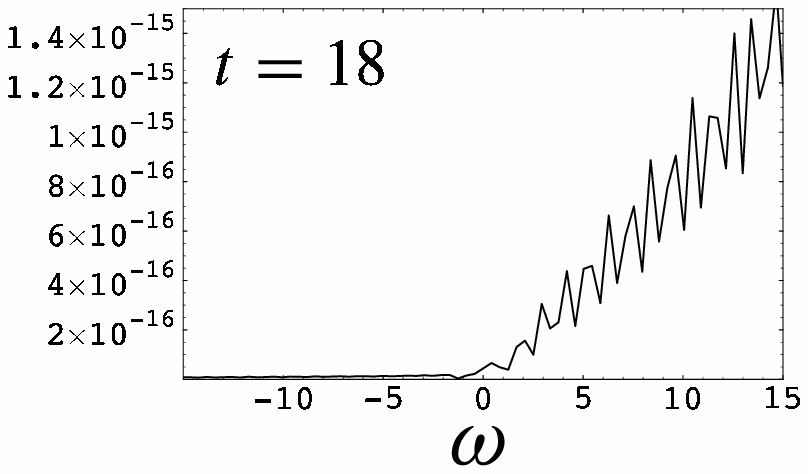}
  \caption{The temporal evolution of the power spectrum
    $\omega^2P_{\text{obs}}(\omega)$ for the %
         output signal in the range $-15\leq\omega\leq 15$.}
   \label{fig:power4}
\end{figure}

The time evolution of the power spectrum of the observed signal is
shown in FIG.~\ref{fig:power2}, FIG.~\ref{fig:power3} and
FIG.~\ref{fig:power4}. The effective frequency $\tilde\omega_0$ of
the input perturbation observed at the inlet of the nozzle is $74.6$.
Figure~\ref{fig:power2} shows that, after $t\sim 10$, the power spectrum
of the observed perturbation spreads out toward the low $\om$ range 
by the redshift effect due to the formation of the sonic horizon, and
the divergence of the power $P_{\text{obs}}(\omega)$ appears at
$\omega=0$. This indicates that the observed spectrum approaches the
theoretically expected form \eqref{eq:power-th} which diverges at
$\omega=0$ as $\propto 1/\omega$. To remove this divergence, we plot
$\omega^2P_{\text{obs}}(\omega)$ in FIG.~\ref{fig:power3} and
FIG.~\ref{fig:power4}.

\subsection{Surface gravity of the sonic horizon}

\begin{figure}[H]
 \centering
 \includegraphics[width=0.33\linewidth,clip]{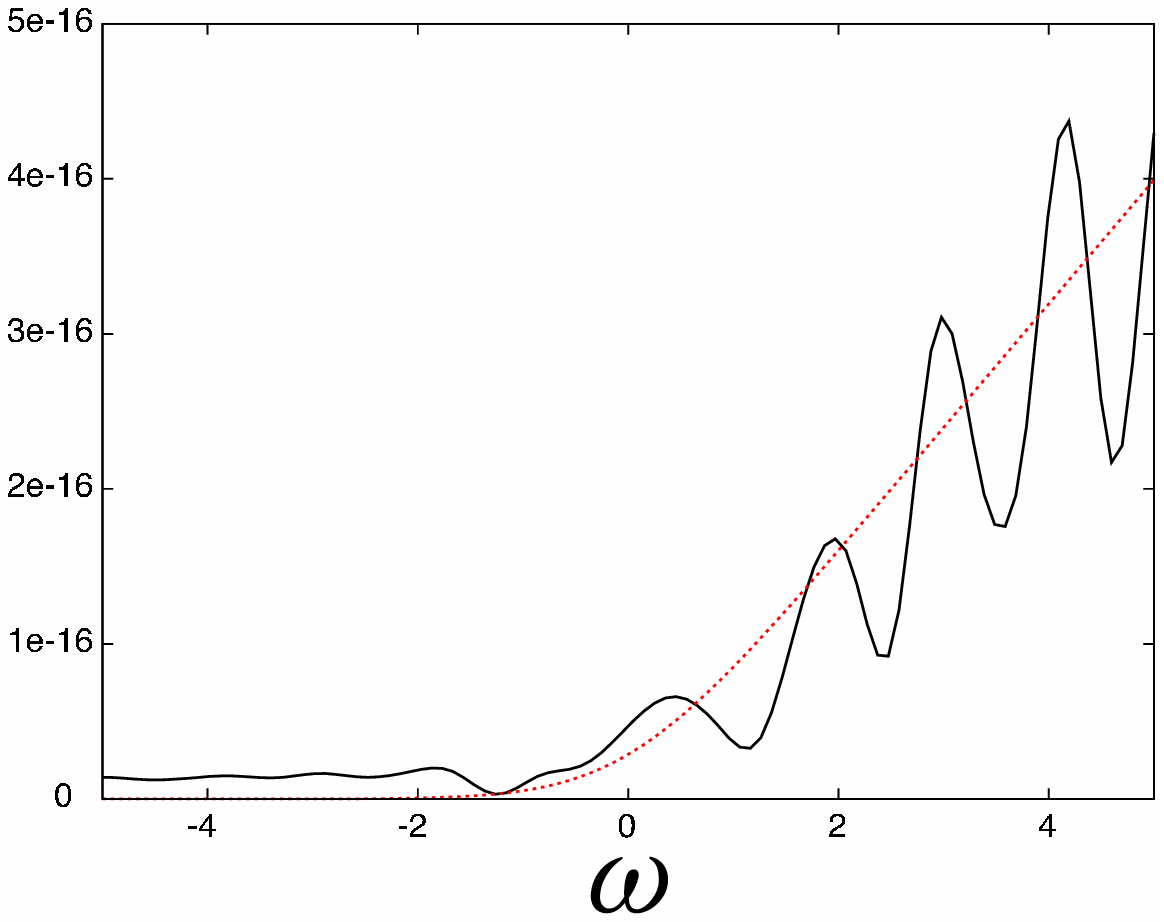}%
 \includegraphics[width=0.33\linewidth,clip]{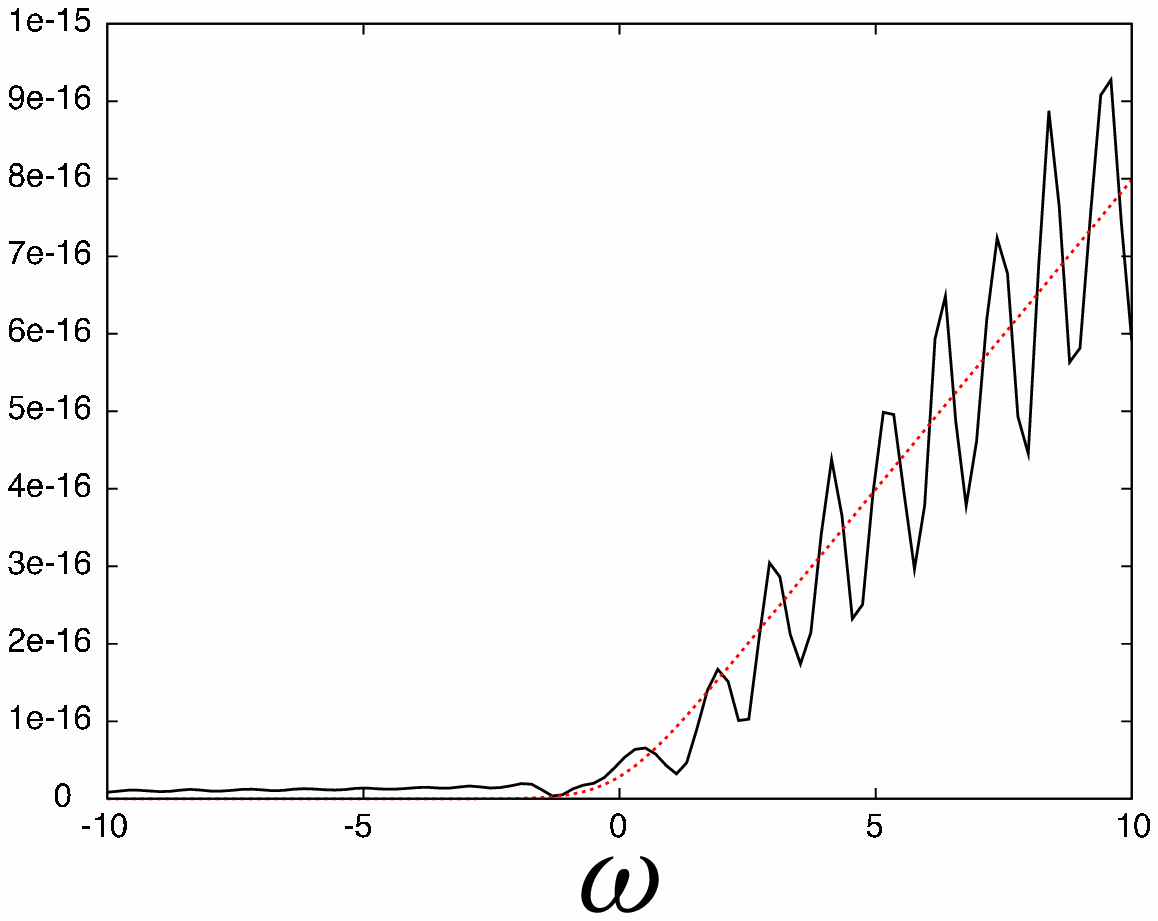}%
 \includegraphics[width=0.33\linewidth,clip]{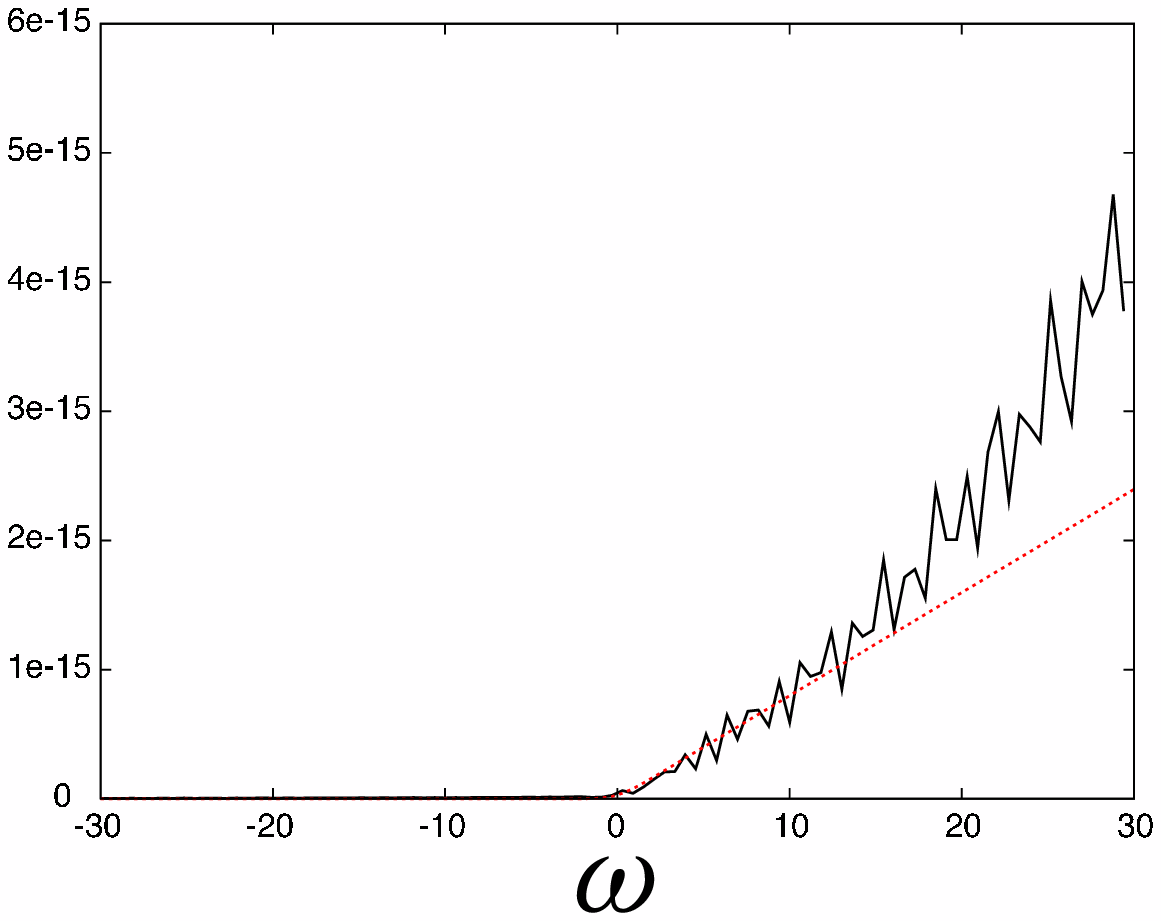}
 \caption{Power spectrums $\om^2 P_{\text{obs}}(\om)$ in different
   ranges of $\omega$. The solid lines are the observed power spectrums
   and the red dotted lines are theoretically expected power
   \eqref{eq:power-th} with $\omega_H=0.360$.}
\label{fig:fit}
\end{figure}

Figure~\ref{fig:fit} shows the observed power spectrum
$\omega^2P_{\text{obs}}(\om)$ at $t=18$ given by Eq.~\eqref{eq:power-exp} and
the theoretically expected power $\omega^2P_{\text{theory}}(\om)$
given by Eq.~\eqref{eq:power-th}.
\noindent
We fits the theoretical power spectrum to the numerical one in
the range $-10\leq\omega\leq 10$ with the parameters
$c_{s1}=1.00, v_1=-0.20, A_J=1.00\times 10^{-6}$ and
$\tilde\omega_0=74.6$, where the value of $\tilde\omega_0$ is read off
from the location of the highest peak of the power in
FIG.~\ref{fig:power3}. Then we obtain the surface gravity of the sonic
horizon
\begin{equation}
  \om_H=\frac{\kappa_H}{2\pi}=0.360\pm 0.013 \,.
\end{equation}
This value is consistent with the theoretically expected value
0.355 given by Eq.~\eqref{eq:temp-th}.

Here we discuss about two points. The first is about the oscillation
of the numerical power spectrum. This oscillation is due to
the finite size Fourier transformation. This oscillation will disappear
if the numerical calculation is carried out with larger value of
$\omega_0$. 

The second point is the deviation of the numerical power from the
theoretical one in the high $\om$ range. This deviation is due to the
input sound wave given in the boundary condition \eqref{eq-ns.bc}. The
input sound wave emerges from the outlet of the nozzle toward the inlet
of the nozzle, and the observation is done at the inlet. Before the
formation of the sonic point, the input sound wave comes from the outlet
to the inlet without receiving a large redshift. Then, as the fluid flow
evolves in time, the redshift on the observed sound wave becomes
larger, and the observed frequency continues to decrease until the
sonic point appears, as denoted by Eq.~\eqref{eq:eff-omega}. Hence,
although the input sound wave is a monochromatic wave at the outlet of
the nozzle (the starting point of the propagation), the observed sound
wave at the inlet of the nozzle has a broad power spectrum after the
formation of the sonic point because of the stimulated effect of the
redshift before the formation of the sonic point. The peak of this broad
spectrum appears obviously in FIG.~\ref{fig:power3} around $\om \sim 75$.
This broadness in the power spectrum is the origin of the deviation of
the numerical power from the theoretical one in FIG.~\ref{fig:fit}.
The perfect thermal spectrum should be realized
at the infinite future, and that the theoretical power spectrum
\eqref{eq:power-th} is the form evaluated at the infinite future.
However we have to make the observation in a finite temporal interval.
Therefore, when we plot the numerical result in an appropriately large
range of $\om$, it is impossible to avoid some deviation of the numerical
power from the theoretical one. If we carry out the numerical calculation
longer and longer, a better agreement have to be obtained in
FIG.~\ref{fig:fit}.

From above discussions, we conclude that FIG.~\ref{fig:fit} shows
the good agreement between the numerically obtained power and
the theoretically expected form of the classical counterpart to
Hawking radiation.

\section{\label{sec:summary}Summary and conclusion}

For the acoustic black hole, the classical counterpart to Hawking
radiation is given by the power spectrum of the perturbation of the
velocity potential of the fluid and detectable in practical experiments.
To demonstrate its detectability, we performed the numerical simulation
of the acoustic black hole in the Laval nozzle and observed the
classical counterpart to Hawking radiation. We obtained the good
agreement of the numerically observed power spectrum with the
theoretically expected one.

Through our numerical simulation, we have obtained two noteworthy
points for data analysis of the experiments of the acoustic black hole:
the first one is that a single input wave can not give us necessary
information of the classical counterpart to Hawking radiation.
We need to carry out several independent observations of the sound
waves with different initial phases to retrieve the thermal distribution
of the power spectrum. The second one is that we can evaluate the
velocity potential at the observation point by integrating the fluid
velocity at that point with respect to time (see
Eq.~\eqref{eq:potential}).  Therefore, it is not necessary to observe
the sound wave at every spatial points in the fluid.

In our calculation presented here, we have considered the sound wave
with small amplitude. However our numerical code is applicable beyond
perturbation, in which the non-linear effect of the sound wave
becomes important. By analyzing such a situation, we may be able to
discuss the backreaction effect on the classical counterpart to Hawking
radiation. As the next step, we are planning to extend our simulation
of transonic flows in a Laval nozzle to include quantum effects. Then,
we expect to obtain implications for quantum aspects of Hawking
radiation using the sonic analogue model of black hole.

\begin{acknowledgments}
We would like to express our gratitude to Satoshi Okuzumi and Masa-aki
Sakagami. Their numerical calculation of the perturbation equation
(not a full order calculation) of the velocity potential was very
impressive for us and we were led to refine our numerical simulation.
\end{acknowledgments}

\begin{thebibliography}{14}
\expandafter\ifx\csname natexlab\endcsname\relax\def\natexlab#1{#1}\fi
\expandafter\ifx\csname bibnamefont\endcsname\relax
  \def\bibnamefont#1{#1}\fi
\expandafter\ifx\csname bibfnamefont\endcsname\relax
  \def\bibfnamefont#1{#1}\fi
\expandafter\ifx\csname citenamefont\endcsname\relax
  \def\citenamefont#1{#1}\fi
\expandafter\ifx\csname url\endcsname\relax
  \def\url#1{\texttt{#1}}\fi
\expandafter\ifx\csname urlprefix\endcsname\relax\def\urlprefix{URL }\fi
\providecommand{\bibinfo}[2]{#2}
\providecommand{\eprint}[2][]{\url{#2}}

\bibitem[{\citenamefont{Hawking}(1975)}]{HawkingSW:CMP43:1975}
\bibinfo{author}{\bibfnamefont{S.~W.} \bibnamefont{Hawking}},
  \bibinfo{journal}{Commu. Math. Phys.} \textbf{\bibinfo{volume}{43}},
  \bibinfo{pages}{199} (\bibinfo{year}{1975}).

\bibitem[{\citenamefont{Birrell and Davis}(1982)}]{BirrellND:CUP:1982}
\bibinfo{author}{\bibfnamefont{N.~D.} \bibnamefont{Birrell}} \bibnamefont{and}
  \bibinfo{author}{\bibfnamefont{P.~C.~W.} \bibnamefont{Davis}},
  \emph{\bibinfo{title}{Quantum fields in curved space}}
  (\bibinfo{publisher}{Cambridge University Press}, \bibinfo{year}{1982}).

\bibitem[{\citenamefont{Visser}(2003)}]{VisserM:IJMPD12:2003}
\bibinfo{author}{\bibfnamefont{M.}~\bibnamefont{Visser}},
  \bibinfo{journal}{Int. J. Mod. Phys. D} \textbf{\bibinfo{volume}{12}},
  \bibinfo{pages}{649} (\bibinfo{year}{2003}).

\bibitem[{\citenamefont{Unruh and Schtzhold}(2005)}]{UnruhW:PRD71:2005}
\bibinfo{author}{\bibfnamefont{W.~G.}~\bibnamefont{Unruh}}, \bibnamefont{and}
  \bibinfo{author}{\bibfnamefont{R.} \bibnamefont{Schutzhold}},
  \bibinfo{journal}{Phys. Rev. D} \textbf{\bibinfo{volume}{71}},
  \bibinfo{pages}{024028} (\bibinfo{year}{2005}).

\bibitem[{\citenamefont{Unruh}(1981)}]{UnruhWG:PRL46}
\bibinfo{author}{\bibfnamefont{W.~G.} \bibnamefont{Unruh}},
  \bibinfo{journal}{Phys. Rev. Lett.} \textbf{\bibinfo{volume}{46}},
  \bibinfo{pages}{1351} (\bibinfo{year}{1981}).

\bibitem[{\citenamefont{Unruh}(1995)}]{UnruhWG:PRD51:1995}
\bibinfo{author}{\bibfnamefont{W.~G.} \bibnamefont{Unruh}},
  \bibinfo{journal}{Phys. Rev. D} \textbf{\bibinfo{volume}{51}},
  \bibinfo{pages}{2827} (\bibinfo{year}{1995}).

\bibitem[{\citenamefont{Visser}(1998)}]{VisserM:CQG15:1998}
\bibinfo{author}{\bibfnamefont{M.}~\bibnamefont{Visser}},
  \bibinfo{journal}{Class. Quantum Grav.} \textbf{\bibinfo{volume}{15}},
  \bibinfo{pages}{1767} (\bibinfo{year}{1998}).

\bibitem[{\citenamefont{Volovik}(1999)}]{VolovikGE:JETPL69:1999}
\bibinfo{author}{\bibfnamefont{G.}~\bibnamefont{Volovik}},
  \bibinfo{journal}{JETP Lett.} \textbf{\bibinfo{volume}{69}},
  \bibinfo{pages}{705} (\bibinfo{year}{1999}).

\bibitem[{\citenamefont{Novello et~al.}(2002)\citenamefont{Novello, Visser, and
  Volovik}}]{NovelloM:WS:2002}
\bibinfo{editor}{\bibfnamefont{M.}~\bibnamefont{Novello}},
  \bibinfo{editor}{\bibfnamefont{M.}~\bibnamefont{Visser}}, \bibnamefont{and}
  \bibinfo{editor}{\bibfnamefont{G.}~\bibnamefont{Volovik}}, eds.,
  \emph{\bibinfo{title}{Artificial Black Holes}} (\bibinfo{publisher}{World
  Scientific}, \bibinfo{year}{2002}).

\bibitem[{\citenamefont{Barcel{\'o} et~al.}(2003)\citenamefont{Barcel{\'o},
  Liberati, and Visser}}]{BarceloC:INJMPA18:2003}
\bibinfo{author}{\bibfnamefont{C.}~\bibnamefont{Barcel{\'o}}},
  \bibinfo{author}{\bibfnamefont{S.}~\bibnamefont{Liberati}}, \bibnamefont{and}
  \bibinfo{author}{\bibfnamefont{M.}~\bibnamefont{Visser}},
  \bibinfo{journal}{Int. J. Mod. Phys. A} \textbf{\bibinfo{volume}{18}},
  \bibinfo{pages}{3735} (\bibinfo{year}{2003}).

\bibitem[{\citenamefont{Barcel{\'o} et~al.}(2004)\citenamefont{Barcel{\'o},
  Liberati, Sonego, and Visser}}]{BarceloC:NJP6:2004}
\bibinfo{author}{\bibfnamefont{C.}~\bibnamefont{Barcel{\'o}}},
  \bibinfo{author}{\bibfnamefont{S.}~\bibnamefont{Liberati}},
  \bibinfo{author}{\bibfnamefont{S.}~\bibnamefont{Sonego}}, \bibnamefont{and}
  \bibinfo{author}{\bibfnamefont{M.}~\bibnamefont{Visser}},
  \bibinfo{journal}{New J. Phys.} \textbf{\bibinfo{volume}{6}},
  \bibinfo{pages}{186} (\bibinfo{year}{2004}).

\bibitem[{\citenamefont{Barcel{\'o} et~al.}(2005)\citenamefont{Barcel{\'o},
  Liberati, and Visser}}]{BarceloC:0505065:2005}
\bibinfo{author}{\bibfnamefont{C.}~\bibnamefont{Barcel{\'o}}},
  \bibinfo{author}{\bibfnamefont{S.}~\bibnamefont{Liberati}}, \bibnamefont{and}
  \bibinfo{author}{\bibfnamefont{M.}~\bibnamefont{Visser}},
  \bibinfo{journal}{Living Rev. Relativity} \textbf{\bibinfo{volume}{8}},
  \bibinfo{pages}{1} (\bibinfo{year}{2005}).

\bibitem[{\citenamefont{Nouri-Zonoz and
  Padmanabhan}(1998)}]{NouriZonozM:9812088:1998}
\bibinfo{author}{\bibfnamefont{M.}~\bibnamefont{Nouri-Zonoz}} \bibnamefont{and}
  \bibinfo{author}{\bibfnamefont{T.}~\bibnamefont{Padmanabhan}},
  \bibinfo{journal}{gr-qc/9812088}  (\bibinfo{year}{1998}).

\bibitem[{\citenamefont{Sakagami and Ohashi}(2002)}]{SakagamiM:PTP107:2002}
\bibinfo{author}{\bibfnamefont{M.}~\bibnamefont{Sakagami}} \bibnamefont{and}
  \bibinfo{author}{\bibfnamefont{A.}~\bibnamefont{Ohashi}},
  \bibinfo{journal}{Prog. Theor. Phys.} \textbf{\bibinfo{volume}{107}},
  \bibinfo{pages}{1267} (\bibinfo{year}{2002}).

\bibitem[{\citenamefont{Yabe et~al.}(2001)\citenamefont{Yabe, Xian, and
  Utusmi}}]{YabeT:JCP169:2001}
\bibinfo{author}{\bibfnamefont{T.}~\bibnamefont{Yabe}},
  \bibinfo{author}{\bibfnamefont{F.}~\bibnamefont{Xian}}, \bibnamefont{and}
  \bibinfo{author}{\bibfnamefont{T.}~\bibnamefont{Utusmi}},
  \bibinfo{journal}{J. Comput. Phys.} \textbf{\bibinfo{volume}{169}},
  \bibinfo{pages}{556} (\bibinfo{year}{2001}).

\bibitem[{\citenamefont{Frolov and Novikov}(1998)}]{FrolovYP:1998}
\bibinfo{author}{\bibfnamefont{Y.~P.} \bibnamefont{Frolov}} \bibnamefont{and}
  \bibinfo{author}{\bibfnamefont{I.~D.} \bibnamefont{Novikov}},
  \emph{\bibinfo{title}{Black Hole Physics}} (\bibinfo{publisher}{Kluwer
  Academic Publishers}, \bibinfo{year}{1998}).

\end{thebibliography}


\appendix*
\section{\label{sec:hawking-radiation}Classical and quantum effects in Hawking radiation}

We briefly review the Hawking radiation in a spacetime of a
gravitational collapse forming a Schwarzschild black hole, and clarify
the distinction between the quantum effects and the classical effects
in the occurrence of the Hawking radiation. For simplicity, we consider
a massless free scalar field $\Phi$ as a representative of the matter
field, and set $c=k_B=1$.

For the first, we start with the classical effects. Because the spacetime
is dynamical, the positive frequency modes and the negative frequency
modes of the scalar field are mixed as the system evolves in time.
This mixing is represented by the Bogoliubov transformation between
the positive frequency mode $\phi_{\om}$ of $\Phi$ at the past null
infinity and that mode $\tphi_{\omega}$ at the future null infinity
\begin{equation}
\phi_{\om} = \int^{\infty}_0 d\omega_1\left[ A_{\om\omega_1}\,
  \tphi_{\omega_1} + B_{\om\omega_1}\, \tphi^{\ast}_{\omega_1} \,\right] \, ,
\label{eq-cc.bogo.trans}
\end{equation}
where $\tphi^{\ast}_{\omega} = \tphi_{-\omega}$ is the negative
frequency mode at the future null infinity. The Bogoliubov
coefficients $A$ and $B$ are obtained  by solving the classical
wave equation $\Box \Phi = 0$ with appropriate boundary conditions at
the past and future null infinities and with the following relation,
\begin{equation}
  \int_0^{\infty} d\omega
\left[A_{\om \om_1}\, A_{\om \om_2}^{\ast} - B_{\om \om_1}\, B_{\om
    \om_2}^{\ast} \right] = \delta_{\om_1\om_2} \,,
\end{equation}
which comes from the normalization condition $(\phi_{\om_1} ,
\phi_{\om_2} ) = \delta_{ \om_1\om_2}$ with respect to the
Klein-Gordon inner product of $\Phi$ \cite{BirrellND:CUP:1982}. This
means that the derivation of $A$ and $B$ is purely classical.

Next, in the classical framework, we consider the wave mode which
propagates from the past null infinity to the future null infinity via
a vicinity of the black hole horizon (see the left panel in
FIG.~\ref{fig:nozzle}). This wave mode is ingoing at the past
null infinity and becomes outgoing at the future null infinity. The
ingoing positive frequency mode $\phi_{\om}$ at the past null
infinity is given by 
\begin{equation}
\phi_{\om}(w) = \frac{1}{\sqrt{4 \pi
    \om}}\exp\left(- i \om  w \right) , \quad \om
> 0 \, ,
\end{equation}
where $w$ is the ingoing null coordinate appropriate for the rest
observer at the past null infinity. When this wave mode
$\phi_{\om}(w)$ propagates to the future null infinity via a vicinity
of the horizon, it evolves to be $\phi_{\om}(w(u))$ at the future
null infinity under the geometrical optics approximation, where $u$
is the outgoing null coordinate appropriate for the rest observer at
the future null infinity. The function $w(u)$ expresses the extremely
large redshift which the mode $\phi_{\om}(w)$ receives during
propagating from the past null infinity to the future null infinity.
This redshift effect can be decomposed into two parts: one is the
redshift during the propagation from the past null infinity to
the vicinity of the horizon, and the other part is the redshift
after passing through the vicinity of the horizon. The first
contribution is not so large and  the mixing of positive and
negative frequency modes does not occur. However, the second
contribution is large enough to make the Bogoliubov coefficient
$B_{\omega\tilde\omega}$ non zero. By matching the null coordinates
$w$ and $u$ along a null geodesic which connects the past and
future null infinities via the vicinity of the horizon,
the function $w(u)$ is obtained
\cite{HawkingSW:CMP43:1975,BirrellND:CUP:1982}:
\begin{equation}
w(u) = - \alpha
\exp\left(- \kappa\, u \right) + \delta,
\label{eq-cc.redshift}
\end{equation}
where $\kappa$ is the surface gravity of the black hole horizon,
$\delta$ is an arbitrary constant denoting the freedom of choosing
the origin of $w$, and the constant $\alpha$ is determined by the first
part of the redshift. The exponential form $\exp(-\kappa u)$ in
the Eq.~\eqref{eq-cc.redshift} comes from the second part of the redshift
and implies that the wave length of the outgoing wave is exponentially
stretched during propagating from a vicinity of the horizon to
the future null infinity. That is, the wave $\phi_{\om}(w(u))$ is
no longer a pure positive frequency mode but becomes a superposition of
positive and negative frequency modes at the future null infinity.
In order to calculate the superposition, we need the outgoing positive
frequency mode $\tphi_{\om}$ at the future null infinity:
\begin{equation}
\tphi_{\om}(u) = \frac{1}{\sqrt{4 \pi \om}}\,
\exp\left(- i \om\,u \right), \quad \om > 0.
\end{equation}
Then, using the definition of the Bogoliubov transformation
\eqref{eq-cc.bogo.trans}, the wave $\phi_{\omega}(w(u))$ at the
future null infinity is decomposed as 
\begin{equation}
 \phi_{\om}(w(u))
   =\int_0^{\infty} d\om_1 \left[A_{\om\om_1}\, \tphi_{\om_1}(u) +
  B_{\om\om_1}\, \tphi^{\ast}_{\om_1}(u)\right].
\label{eq-cc.mode.future}
\end{equation}
The Bogoliubov coefficients are obtained using the inner product  $(
\phi_{\om_1} , \tphi_{\om_2} ) = A_{\om_1\om_2}$ and $(\phi_{\om_1} ,
\tphi^{\ast}_{\om_2} ) =-B_{\om_1\om_2}$, and we find
\begin{equation}
\left|A_{\om_1\om_2}\right|^2 = e^{\om_2/\om_H} \left|B_{\om_1
    \om_2} \right|^2 , \quad 
\left| B_{\om_1\om_2} \right|^2 =
\frac{1}{2\pi\kappa\om_1}\frac{1}{e^{\om_2/\om_H} - 1}, \quad
 \om_H = \frac{\kappa}{2 \pi} .
\label{eq-cc.bogo.coeff}
\end{equation}
The square of the Bogoliubov coefficients do not depend on the
constants $\alpha$ and $\delta$, and the following relation holds:
\begin{equation}
\left| B_{\om_1, -\om_2} \right|^2 =
\left| A_{\om_1\om_2} \right|^2 \, .
\label{eq-cc.bogo.coeff.2}
\end{equation}
All the above phenomena are the classical effects.

Finally we proceed to the quantum effects. When $\Phi$ is quantized,
the harmonic operators $a_{\om}$ and $a_{\om}^{\dag}$ with respect to
the past mode $\phi_{\om}$ are related to those $\ta_{\om}$ and
$\ta_{\om}^{\dag}$ with respect to the future mode $\tphi_{\om}$
as follows\cite{HawkingSW:CMP43:1975,BirrellND:CUP:1982}:
\begin{align}
&\ta_{\om} = \int_0^{\infty} d\om_1\left( A_{\om_1\om}\, a_{\om_1} +
  B_{\om_1\om}^{\ast}\, a_{\om_1}^{\dag} \right)  , \quad
\ta_{\om}^{\dag} = \int_0^{\infty} d\om_1\left( A_{\om_1
    \om}^{\ast}\, a_{\om_1}^{\dag} + B_{\om_1\om}\, a_{\om_1}\right), \\
&[a_{\omega_1},a^{\dagger}_{\omega_2}]=\hbar\,\delta_{\omega_1\omega_2},\quad
[\tilde a_{\omega_1},\tilde a^{\dagger}_{\omega_2}]=\hbar\,\delta_{\omega_1\omega_2}.
\end{align}
Then the number of particles at the future null infinity is
obtained
\begin{equation}
 N_{\om} = \left<0\right|\ta^{\dag}_{\om} \, \ta_{\om}\left|0\right>
 = \hbar\int^{\infty}_0d\om_1\left|B_{\om_1\om} \right|^2 \, ,
\label{eq-cc.qft}
\end{equation}
where $\left|0\right>$ is the vacuum state at the past null infinity,
$a_{\om}\left|0\right> = 0$. Therefore, for the mode $\phi_{\om}$ which
passes a vicinity of the horizon and propagates to the future null
infinity, using Eqs.~\eqref{eq-cc.bogo.coeff} and an appropriate
regularization method \cite{HawkingSW:CMP43:1975}, we obtain
\begin{equation}
 N_{\om} = \left<0\right| \, \ta^{\dag}_{\om}
\, \ta_{\om} \, \left|0\right> = \frac{\hbar}{e^{\hbar\om/\hbar\om_H} - 1} \, .
\label{eq-cc.hr}
\end{equation}
It is concluded that a black hole emits a thermal radiation of $\Phi$
with the Hawking temperature $T_H \equiv \hbar\, \om_H$. 

It should be emphasized that, although the creation of particles is just
the quantum effect, however the Planckian distribution (\ref{eq-cc.hr})
of the emitted particles is purely the classical effect due to
Eqs.~\eqref{eq-cc.bogo.coeff}. The thermal nature of the spectrum comes
from the Bogoliubov coefficient $|B_{\om_1\om_2}|^2$ which has the
Planckian distribution with respect to $\om_2$. That is, the thermal
nature of the Hawking radiation comes from the extremely large redshift
for the wave mode $\phi_{\om}$ which passes the vicinity of the horizon.
If we take the classical limit $\hbar\rightarrow 0$, the number of
created particles and the Hawking temperature become zero but $|B|^2$
is not affected.

\end{document}